\documentclass[10pt,journal,compsoc]{IEEEtran}
\usepackage{times, alltt, xspace}
\usepackage{amsmath}
\usepackage{amssymb}
\usepackage{setspace}
\usepackage{graphicx}
\usepackage{tabularx}
\usepackage{algpseudocode}
\usepackage{algorithm}
\usepackage{multirow}
\usepackage{caption}
\usepackage{subcaption}
\usepackage{etoolbox}
\usepackage{enumitem}
\usepackage{booktabs}
\usepackage{multicol}
\usepackage{tabularx}
\usepackage{epstopdf}
\makeatletter
\patchcmd{\@begintheorem}{\textit}{\textbf}{}{}
\makeatother

\usepackage{amsthm}
\newtheoremstyle{break}
  {}
  {}
  {}
  {}
  {\bfseries}
  {.}
  {4pt}
  {}
\theoremstyle{break}
\makeatletter
\long\def\@IEEEtitleabstractindextextbox#1{\parbox{0.922\textwidth}{#1}}
\makeatother

\newcommand{\U}{\ensuremath{\mathrm{U}}\xspace}
\newcommand{\UG}{\ensuremath{\mathrm{UG}}\xspace}
\newcommand{\ATTR}{\ensuremath{\mathrm{UA}}\xspace}

\newcommand{\miniGURA}{\ensuremath{\mathrm{rGURA}}}
\newcommand{\hgabac}{\ensuremath{\mathrm{HGABAC}}\xspace}

\newcommand{\SR}{\ensuremath{\mathrm{SR_d}}}

\newcommand{\Range}{\ensuremath{\mathrm{Range}}}

\newcommand{\miniGURAN}{\ensuremath{\mathrm{rGURA}}}
\newcommand{\miniGURANZ}{\ensuremath{\mathrm{rGURA_0}}}
\newcommand{\miniGURANO}{\ensuremath{\mathrm{rGURA_1}}}
\newcommand{\miniGURANG}{\ensuremath{\mathrm{rGURA_G}}\xspace}
\newcommand{\miniGURANGZ}{\ensuremath{\mathrm{rGURA_{G_0}}}\xspace}
\newcommand{\miniGURANGZN}{\ensuremath{\mathrm{rGURA_{G_0}}}}

\newcommand{\miniGURANGZB}{\ensuremath{\mathrm{\mathbf{rGURA_{G_0}}}}\xspace}
\newcommand{\miniGURANGO}{\ensuremath{\mathrm{rGURA_{G_1}}}\xspace}
\newcommand{\miniGURANGOP}{\ensuremath{\mathrm{rGURA_{G_{1+}}}}\xspace}
\newcommand{\miniGURANGT}{\ensuremath{\mathrm{rGURA_{G_{1+}}}}\xspace}
\newcommand{\miniGURANGOB}{\ensuremath{\mathrm{\mathbf{rGURA_{G_1}}}}\xspace}
\newcommand{\miniGURANGTB}{\ensuremath{\mathrm{\mathbf{rGURA_{G_{1+}}}}}\xspace}
\newcommand{\canA}{{\mathsf{canAddU}}}
\newcommand{\canD}{{\mathsf{canDeleteU}}}
\newcommand{\canAUG}{{\mathsf{canAddUG}}}
\newcommand{\canDUG}{{\mathsf{canDeleteUG}}}
\newcommand{\Assign}{{\mathsf{canAssign}}}
\newcommand{\Remove}{{\mathsf{canRemove}}}

\newcommand{\GURA}{\ensuremath{\mathrm{GURA}}\xspace}

\newcommand{\rGURA}{\ensuremath{\mathrm{\miniGURAN}}\xspace}
\newcommand{\GURAG}{\ensuremath{\mathrm{GURA_G}}\xspace}
\newcommand{\rGURAG}{\ensuremath{\mathrm{rGURA_G}}\xspace}
\newcommand{\rGURAGN}{\ensuremath{\mathrm{rGURA_G}}}

\newcommand{\effu}{\ensuremath{\mathrm{effU}}}
\newcommand{\effug}{\ensuremath{\mathrm{effUG}}}
\newcommand{\directug}{\ensuremath{\mathrm{directUg}}}

\newcommand{\effectiveug}{\ensuremath{\mathrm{effUg}}}

\newcommand{\ps}{\ensuremath{\mathrm{PSPACE}}}

\algnewcommand{\LeftComment}[1]{\Statex \(\triangleright\) #1}

\newcommand{\N}{$\mathrm{\overline{N}}$\xspace}
\newcommand{\D}{$\mathrm{\overline{D}}$\xspace}

\newcommand{\I}{\ensuremath{\mathrm{I}}\xspace}
\newcommand{\A}{\ensuremath{\mathrm{SUBAR}}}

\newcommand{\RPSUPER}{\ensuremath{\mathrm{RP_{\supseteq}}}\xspace}
\newcommand{\RPSAME}{\ensuremath{\mathrm{RP_{=}}}\xspace}

\newcommand{\REQ}{\ensuremath{\mathrm{REQ}}\xspace}
\newcommand{\AR}{\ensuremath{\mathrm{AR}}\xspace}

\newcommand{\SUBAR}{\ensuremath{\mathrm{SUBAR}}}

\newcommand{\C}{\ensuremath{\mathrm{C}}}
\newcommand{\Q}{\ensuremath{\mathrm{Q}}}
\newcommand{\SCOPE}{\ensuremath{\mathrm{SCOPE}}}

\newcommand{\UAAN}{\ensuremath{\mathrm{UAA}}}

\newcommand{\effatt}{\ensuremath{e\_{att}}}
\newcommand{\UGAAN}{\ensuremath{\mathrm{UGAA}}}

\newcommand{\UGAN}{\ensuremath{\mathrm{UGA}}}

\newcommand{\false}{\ensuremath{\mathrm{\mathbf{false}}}}
\newcommand{\true}{\ensuremath{\mathrm{\mathbf{true}}}}
\newcommand{\Evaluate}{\ensuremath{\mathrm{Satisfy}}}
\newcommand{\?}[1]{\kern-.#1em }
\newtheorem{thm}{Theorem}
\newtheorem{col}{Corollary}
\newtheorem{lem}{Lemma}
\newtheorem{mydef}{Definition}

\ifCLASSOPTIONcompsoc
  \usepackage[nocompress]{cite}
\else
  \usepackage{cite}
\fi

\ifCLASSINFOpdf

\else

\fi

\begin{document}
\title{Reachability Analysis for Attributes in \\ ABAC with Group Hierarchy}

\author{Maanak~Gupta,~\IEEEmembership{Member,~IEEE}
        and~Ravi~Sandhu,~\IEEEmembership{Fellow,~IEEE}
\IEEEcompsocitemizethanks{\IEEEcompsocthanksitem Maanak Gupta is with the Department
of Computer Science, Tennessee Tech University, Cookeville, TN, 38501, USA.
E-mail: mgupta@tntech.edu}
\IEEEcompsocitemizethanks{\IEEEcompsocthanksitem Ravi Sandhu is with the Institute for Cyber Security and Department of Computer Science, University of Texas at San Antonio, San Antonio, TX, 78249, USA.
E-mail: ravi.sandhu@utsa.edu}}

\markboth{Gupta and Sandhu: Reachability Analysis for Attributes in ABAC with Group Hierarchy}%
{Shell \MakeLowercase{\textit{et al.}}: Bare Demo of IEEEtran.cls for Computer Society Journals}

\IEEEtitleabstractindextext{%
\begin{abstract}
Attribute-based access control (ABAC) models are widely used to provide fine-grained and adaptable authorization based on the attributes of users, resources, and other relevant entities. Hierarchial group and attribute based access control (HGABAC) model was recently proposed which introduces the novel notion of attribute inheritance through group membership. GURA\textsubscript{G} was subsequently proposed to provide an administrative model for user attributes in HGABAC, building upon the ARBAC97 and GURA administrative models.  The GURA model uses administrative roles to manage user attributes.  The reachability problem for the GURA model is to determine what attributes a particular user can acquire, given a predefined set of administrative rules. This problem has been previously analyzed in the literature.  In this paper, we study the user attribute reachability problem based on directly assigned attributes of the user and attributes inherited via group memberships. We first define a restricted form of GURA\textsubscript{G}, called rGURA\textsubscript{G} scheme, as a state transition system with multiple instances having different preconditions and provide reachability analysis for each of these schemes. In general, we show PSPACE-complete complexity for all rGURA\textsubscript{G} schemes. We further present polynomial time algorithms to solve special instances of rGURA\textsubscript{G} schemes under restricted conditions.
\end{abstract}

}

\maketitle

\IEEEdisplaynontitleabstractindextext
\IEEEpeerreviewmaketitle
\IEEEraisesectionheading{\section{Introduction}\label{sec:introduction}}
\IEEEPARstart{A}{ttribute}-based access control (ABAC) is considered as an important authorization system among practitioners and researchers. The system offers fine-grained and adaptable access control solutions based on the characteristics, referred to as attributes, of several entities. ABAC systems provide a flexible and scalable approach to secure resources in distributed environments and overcome some of the shortcomings of traditional discretionary access control (DAC)\cite{sandhu1994access}, mandatory access control (MAC) \cite{sandhu1993lattice} and role based access control (RBAC) \cite{sandhu1996role} models.  Several attribute based access control models have been formulated \cite{jin2012unified, wang2004logic, shen2006attribute, hu2014guide, hu2015attribute, goyal2006attribute, lang2009flexible, yuan2005attributed, frikken2006attribute, DBLP:journals/tdsc/TripunitaraL13, cirio2007role,covington2006contextual, gupta2018authorization, gupta2018attribute, gupta2019dynamic, gupta2020attribute, gupta2020secure} but a strong consensus on its definitive characteristics is still to be achieved. More recently, hierarchial group and attribute based access control model (\hgabac) \cite{servos2014hgabac} was proposed, which introduced the notion of user and object groups to assign attributes to users and objects respectively. In this model, besides the direct assignment of attributes to users or objects, groups are also assigned attributes, which are then assigned to users and objects through corresponding group memberships. The most important advantage of this model is the ease of administration, since multiple attributes can be assigned or removed from users or objects through single administrative operation. The administrative model for HGABAC, referred as \GURAG, was defined in \cite{gupta2016mathrm}, to control user attribute assignment based on specified precondition rules and administrative roles. This model has three sub-models UAA (user attribute assignment), UGAA (user group attribute assignment) and UGA (user to user-group assignment) which assigns attributes to users directly or indirectly through groups. The model extends well-known ARBAC97 \cite{sandhu1999arbac97} administrative model and recently published GURA administrative model \cite{jin2012role} by introducing administration of attributes for user-groups and managing user to groups memberships.

 In ABAC, the attributes of an entity are critical in determining its permissions.  Therefore, it is an important question to compute the attribute values that an entity can acquire through the combination of administrative roles and rules. In the context of \GURAG, it is imperative to understand the set of attribute values a user can get based on direct assignment or via group memberships. Group hierarchy also exists in the HGABAC operational model which further complicates computation of the possible effective attribute values of a user. Although security administrators are trusted to assign attributes correctly, it is still desirable to understand the eventual set of attribute values that a user can acquire through multiple direct and indirect assignments. Such analysis can also help to identify a sequence of administrative actions required by administrators to assign certain attribute values to the users. It further allows administrators to know the future attribute values an entity can achieve based on predefined administrative rules, which can help them to understand if certain permissions can ever be granted to an entity.

As the number of attributes, attribute values and administrative rules become large, certain anomalies become hard to detect just by simple inspection. For example, suppose an administrative user having role RoomAdmin is allowed to add user attribute roomAcc with value 1.02 to a user only if the user's attribute status has value Grad and the user does not currently have roomAcc 2.01. Further, a user can be assigned status attribute with value Grad only through group G$_1$ membership since there is no direct assignment administrative rule for status attribute. Another administrative rule assigns roomAcc 2.01 to a group junior to G$_1$, thereby getting G$_1$ with roomAcc value 2.01. Now if a user is assigned to user-group G$_1$, she will get all G$_1$'s attributes, including roomAcc with value 2.01. It might seem that user will not be able to get roomAcc value 1.02 and 2.01 together. However, it is possible if the junior group to G$_1$ is assigned value 2.01 after the user is assigned to user-group G$_1$. Such security policy anomalies can be discovered with the help of reachability analysis, which checks if entities can get certain values together or whether entity will get particular values based on the set of administrative rules defined through administrative models.

In this paper we analyze the attribute reachability analysis focusing on the effective attributes of the user achieved through direct assignment and through user-group memberships. This work extends the reachability analysis \cite{jin2013reachability} done for \GURA~administrative model \cite{jin2012role}, where the attributes were only directly assigned to users without the concept of group memberships. In our analysis, we have defined a restricted \GURAG model, called \rGURAG, which considers a subset of preconditions which can be created in \GURAG. We abstract \rGURAG into a state transition system and specify three separate instances---\miniGURANGZ, \miniGURANGO and \miniGURANGOP---to cover different set of prerequisite conditions for attributes assignments to a user or a group, and also for user to group membership assignment. Our reachability analysis primarily focuses on the effective set of attributes of users which is the union of direct attributes and attributes attained by group membership. We have defined reachability queries which is the required set of effective attributes a user can achieve in any target state. Two different types of reachability queries are discussed, one with the exact values and another with the superset of attribute values. We will show that the general reachability problem for \rGURAG schemes is PSPACE-complete. We further identify certain more restricted cases of \rGURAG schemes where the reachability problem can be solved in polynomial time. For such instances we will provide algorithms and a sequence of administrative requests (referred as reachability plan) to satisfy the reachability query.

The rest of this paper is organized as follows. Section \ref{s:related} reviews the related work. In Section \ref{s:background}, we review the \hgabac model and \GURAG administrative model. Section \ref{s:restricted} discusses the generalized restricted \rGURAG scheme and its instances. In Section \ref{s:problem}, we formally define our user attribute reachability problem. Formal proofs for general \rGURAG schemes are discussed in Section \ref{sec:pspace}. Section \ref{s:proof} presents polynomial algorithms for some restricted versions of \rGURAG schemes followed by example problems instances in Section \ref{sec:example}. Section \ref{s:conc} concludes this paper.

\section{Related Work}
\label{s:related}
Reachability analysis for user attributes was first studied by Jin et al~\cite{jin2013reachability}, based on the GURA administrative model \cite{jin2012role}. In this analysis, attribute values are assigned to users directly based on certain attribute-based prerequisite conditions and by administrators assuming roles. This work proves PSPACE-complete complexity for generalized GURA scheme and also presents polynomial algorithms for some conditional cases. Our work extends the aforementioned reachability analysis where attributes are assigned to users as well as to groups to which users are members. This assignment of attributes to groups provides administrative benefits in addition and removal of multiple attributes to users with a single administrative operation.

Security policies have been widely analysed in several works including \cite{harrison1976protection,tripunitara2007theory, li2006security, li2005beyond, sandhu1988schematic, sasturkar2006policy,sandhu1992typed, lipton1977linear, schaad2002lightweight, jha2008towards, rajkumar2016safety}. The safety analysis problem goes back to 1970’s. In general, the safety of access control matrix (ACM) model was shown to be undecidable in \cite{harrison1976protection}. Tripunitara and Li presented an important theoretical comparison of expressive powers of different access control models in \cite{tripunitara2007theory}. Many of our notations in this paper are adapted from this work. Same authors in \cite{li2006security} defined restricted forms of ARBAC97 (AATU and AAR) and provided algorithms for analysis problems including safety and availability in restricted forms. This works extends results from trust management policies in \cite{li2005beyond} where safety and availability security analysis on delegation of authority is discussed. The schematic protection model (SPM) \cite{sandhu1988schematic} introduced typed security entities where each entity is associated with a security type, which remains unchanged. Sasturkar et al \cite{sasturkar2006policy} analyse ARBAC97 administrative policies to determine reachability and availability problems, by establishing connections between artificial intelligence planning problem. Jha et al \cite{jha2008towards} classified analysis problems related to RBAC and claimed PSPACE-complete solutions for unrestricted classes whereas NP-complete and polynomial time algorithms for restricted subclasses. Lipton et al \cite{lipton1977linear} presented a linear time algorithm for take and grant system. Alloy language is used for specification of role based system and analysis is done using Alloy constraint analyser in \cite{schaad2002lightweight}. Recently, Rajkumar and Sandhu discussed safety problem for pre-authorization sub-model for $\mathrm{UCON_{ABC}}$ in \cite{rajkumar2016safety}.

Jajodia et al \cite{jajodia1997logical} presented a logical language to express positive, negative and derived authorization policies, and provided polynomial algorithms to check completeness and consistency. Cholvy and Cuppens \cite{cholvy1997analyzing} discussed the problem of policy consistency and offered a methodology to solve it. They further suggested the use of roles priorities to resolve normative conflicts in policies. \cite{bandara2003using} provides a method to transform policy specifications into event calculus based formal notation. It further describes the use of abductive logical reasoning to perform a priori analysis of various policy specifications. Jaeger et al \cite{jaeger2003policy} presented the concept of access control space and its use in managing access control policies. These spaces are used to represent permission assignment to subjects or roles. Authors in \cite{fisler2005verification} presented decision diagram based algorithms to analyze XACML based policies and compute the semantic differencing information between versions of policies. Stoller et al \cite{stoller2007efficient} provided algorithms for ARBAC97 policies limited to rules with one positive precondition and unconditional role revocations. Same authors in \cite{stoller2011symbolic} defined PARBAC (parameterized ARBAC) and determined user-reachability problem as undecidable over an infinite types of parameter. It further assumed all parameters as atomic-valued and are changed when the role is modified. Gupta et al \cite{gupta2014abductive} discussed rule-based administrative model to control addition and removal of facts (attributes) and rules. It further proposed an abductive algorithm which can analyse policies even when the facts (attributes) are unavailable based on computation of minimal sets of facts. The work in \cite{joshi2005analysis} provides analysis of expressive power of generalized temporal role-based access control (GTRBAC) which offers a set temporal constraints to specify fine grained time based policies.

Several works \cite{gupta2016mathrm}, \cite{sandhu1999arbac97}, \cite{jin2012role}, \cite{crampton2003administrative} have been presented to discuss administrative models for well known access control models. ARBAC97 \cite{sandhu1999arbac97} discusses the user to role assignment based on the administrative rules comprising of administrative roles and prerequisite conditions based on roles. The \GURAG administrative model \cite{gupta2016mathrm} provides a generalized administrative model for attributes based access control models by asserting role as one of the several user attributes. These works define attribute based preconditions and administrative roles to assign and remove attributes from users and groups. Crampton and Loizou \cite{crampton2003administrative} also presented an administrative work related to RBAC model and developed models for role hierarchy administration.


\section{background}
\label{s:background}
In this section, we will provide an overview of reformalized hierarchial group and attribute based access control (HGABAC) model. We will further discuss the \GURAG model \cite{gupta2016mathrm} and its three sub models user attribute assignment (UAA), user-group attribute assignment (UGAA) and user to user-group assignment (UGA). The main objective of this section is to lay the foundation of our reachability analysis and make the reader familiar with relevant terminologies and concepts.
\subsection{HGABAC Model}
This subsection discusses the reformalized HGABAC model as defined in \cite{gupta2016mathrm}. We have formulated this model in style of $\mathrm{ABAC_\alpha}$ \cite{jin2012unified} to help in our administrative model and reachability analysis. The model is notationally different but equivalent to HGABAC model provided by Servos et al \cite{servos2014hgabac}. We begin with an informal overview of the model followed by formal definitions of components of HGABAC relevant to our reachability analysis.

\subsubsection{Model Overview}
\begin{figure}[t]
\centering
\includegraphics[scale=.50]{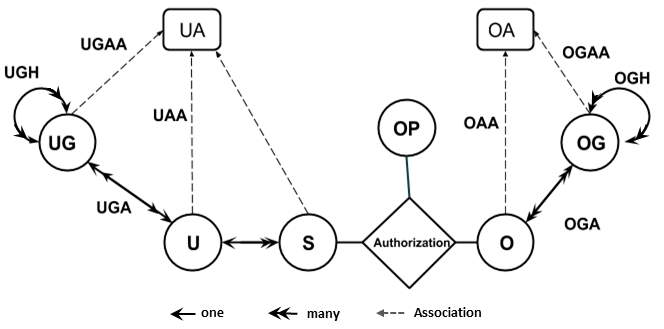}
\caption{HGABAC Conceptual Model}
\label{fig:hgabac}
\end{figure} 
Figure \ref{fig:hgabac} shows the conceptual HGABAC model. The basic components include traditional access control entities like Users (U), Objects (O), and Subjects (S). A user is a human being interacting directly with a computer whereas subject is an active entity (like an application or a process) created by the user to access resources or objects. A user can create multiple subjects but each subject must belong to a single user. OP represents the set of operations which can be performed by subjects on objects. The novel approach introduced by HGABAC model is the notion of user groups (UG) and object groups (OG), which are a collection of users or objects respectively. The set of user and object attributes is defined by UA and OA respectively. Each attribute in set UA and OA is a set-valued function, which takes different entities, like users, objects, user-groups or object-groups, and return values from the attribute range. As the attributes are assigned to groups also, the prime advantage of this assignment is the inheritance of attributes to the group's user or object members. For example, if a user-group $ug$ with attribute skills having values c and java, is assigned to user $u$, then $u$ will inherit attribute skill with values c and java from $ug$. Group hierarchy also exists in HGABAC (defined using a partial relation and shown as self loops in Figure \ref{fig:hgabac}) where senior groups inherit all the attributes from their junior groups. For example, suppose a junior group to $ug$, say $ug'$, is assigned value c++ for attribute skill, then $ug$ will inherit this value and its effective values for skill will be c, java and c++. In this case user $u$ already assigned to user group $ug$ will get all three values for skill attribute. Similar assignments can be done for object and object groups also. It should be noted that each user or object can be assigned to multiple user or object groups and vice versa. A subject inherits all or subset of the effective attributes of the creator user. Each operation $op$ $\in$ OP will have an associated boolean authorization function which specifies the policies under which a subject is allowed to perform operation $op$ on the objects. These policies are specified as propositional logic formulas using the model's policy language and are defined by the security architect at the time of system creation. A subject is allowed to perform operations on an object if the effective attributes of subjects and objects satisfy the boolean authorization function.

\subsubsection{Formal Definitions (User Attributes Only)}\label{ss:hgabac}
\begin{figure}[t]
\centering
\includegraphics[width=.5\textwidth, height=5cm]{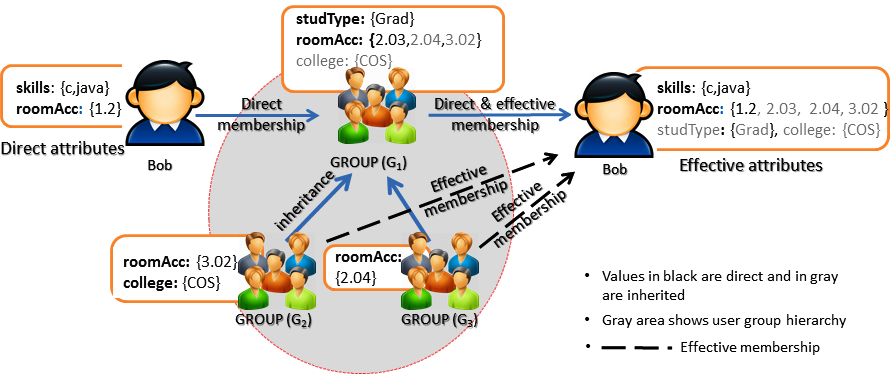}
\caption{ Example User and User Group Attributes
\label{fig:fig example}}
\end{figure}

The \GURAG administrative model \cite{gupta2016mathrm} deals with the user side of HGABAC model reflecting the administrative relations for users and user groups to modify their attributes. Similar administrative model can also be extended for objects but is out of the scope of the paper. Our reachability analysis also considers only the effective attributes of the user, and therefore, we will only formalize the relevant sets, relations and functions pertinent to HGABAC and required in our analysis. Table \ref{tab:hgabac-model} defines the formal HGABAC model covering the required definitions. An example configuration with respect to these definitions is shown in Figure \ref{fig:fig example} and Table \ref{tab:hgabac}.
\begin{table}[]
\centering
\caption{HGABAC Formal Model (User Attributes Only)}
\label{tab:hgabac-model}
\resizebox{\columnwidth}{!}{%
\begin{tabular}{llllll}
\toprule
\multicolumn{6}{l}{\textbf{Basic Sets and Relations}}\\
\multicolumn{6}{l}{-- U, S (finite sets of users and subjects respectively).}\\
\multicolumn{6}{l}{-- UG (finite set of user groups).}\\
\multicolumn{6}{l}{-- UA (finite set of user attribute functions).}\\
\multicolumn{6}{l}{-- For each $att \in$ \ATTR, \SCOPE$_{att}$ is a finite set of atomic values and}\\
\multicolumn{6}{l}{\;\;\;$\Range(att) = \mathcal{P}(\SCOPE_{att})$ where $\mathcal{P}$ denotes the powerset.}\\
\multicolumn{6}{l}{-- UGH $\subseteq$ UG $\times$ UG,  a partial order relation  $\succeq_{ug}$ on UG.}                                                                                                          \\
\multicolumn{6}{l}{ \textbf{Defined (Direct) Functions}}                                                                                                          \\
\multicolumn{6}{l}{\begin{tabular}[c]{@{}l@{}}-- For each $att$ $\in$ \ATTR.~$att$: U $\cup$ UG $\rightarrow $ \Range($att$), maps \\ \;\; each user and user group to a subset of values in \SCOPE$_{att}.$\end{tabular}}
\\
\multicolumn{6}{l}{-- directUg : $ \mathrm{U \rightarrow 2^{UG}} $,  maps each user to a subset of user groups in UG.}                                                                                                          \\
\multicolumn{6}{l}{ \textbf{Derived (Effective) Functions}}                                                                                                          \\
\multicolumn{6}{l}{\begin{tabular}[c]{@{}l@{}}-- $\effectiveug$ : \U $\rightarrow 2^{\UG}$, defined as \\ \;\;\;$\mathrm {\effectiveug(u) = \textit{\directug}(u)} $ $\cup $ \indent \indent \indent \indent $ (\mathrm{\bigcup\limits_{ug_i \; \in\; \directug(u)}  \{{ug_j\;|\; ug_i \succeq_{\it{ug}} ug_j\})}}.$\end{tabular}} \\

\multicolumn{6}{l}{-- For each $att \in$ \ATTR,}                                                                                                          \\
\multicolumn{6}{l}{\begin{tabular}[c]{@{}l@{}} \;\;\;\; $\bullet $ $\effug_{\it{att}}$ : UG $\rightarrow $\Range($att$), defined as \\ \;\;\;\;\;\;\; $\mathrm {\effug_{\it{att}}(ug_i) = {\it{att}} (ug_i)} $ $\cup $ $ (\mathrm{\bigcup\limits_{ g \; \in\; \{ug_j\;|\; ug_i \; \succeq_{\textit{ug}}\; ug_j\} } {\effug}_{\it{att}} (g))}.$\end{tabular}} \\[\defaultaddspace]

\multicolumn{6}{l}{\begin{tabular}[c]{@{}l@{}} \;\;\;\; $\bullet$ $\effu_{\it{att}}$ : \U $\rightarrow$ $\Range(att)$, defined as \\ \;\;\;\;\;\; $\mathrm {\effu_{\it{att}}(u) = {\it{att}} (u) }$ $\cup $ $ (\mathrm{\bigcup\limits_{g \; \in\; \directug(u) } {\effug}_{\it{att}} (g))}.$\end{tabular}} \\
\bottomrule

\end{tabular}
}
\end{table} 

Basic sets and relations as shown in Table \ref{tab:hgabac-model} include \U, S and \UG representing the set of users, subjects and user groups in the system. \ATTR represents the set of user attribute functions for user and user groups where each attribute function in \ATTR is set valued. These attribute functions can assign values to user or user groups from the set of atomic values, represented as \SCOPE$_{att}$. The power set of \SCOPE$_{att}$ is defined by \Range($att$). Example definitions for these sets is shown in first part of Table \ref{tab:hgabac}. User group hierarchy (UGH) is a partial order relation on UG, defined as $\succeq_{ug}$, where $ug_1$ $\succeq_{ug}$ $ug_2$ represents $ug_1$ is senior to $ug_2$ or $ug_2$ is junior to $ug_1$. As shown in gray the area of Figure \ref{fig:fig example}, UGH = $\mathrm{\{(G_1, G_1), (G_2, G_2), (G_3, G_3), (G_1, G_2), (G_1, G_3)\}}$. This UGH relation results in inheritance of attributes from junior to senior group (will discuss in a moment).

Attribute values can be directly assigned to user and user groups which is denoted by function $att$ in \ATTR. As defined in Figure \ref{fig:fig example} and Table \ref{tab:hgabac}, user Bob is directly assigned \{c, java\} for attribute function skills. Similarly, other direct attributes are given for Bob and user groups $\mathrm{G_1, G_2, G_3}$. The function \directug~specifies the user groups to which the user is directly assigned. In our example, Bob is directly assigned to user group $\mathrm{G_1}$. We also define the effective user groups of the user (denoted by \effectiveug), which states all the groups to which the user is either directly or indirectly assigned via UGH relation. Effective user group for Bob will be $\mathrm{\{G_1, G_2, G_3\}}$, since Bob is directly assigned to $\mathrm{G_1}$ and $\mathrm{G_1}$ has junior groups as $\mathrm{G_2}$ and $\mathrm{G_3}$.

The effective values of an attribute $att$ (\effug$_{att}$) for a user group is the union of the user's direct attribute values and the effective values of all its junior groups in UGH relation. Note that this definition is well formed since $\succeq_{\it{ug}}$ is a partial order. For the minimal groups $ug_j$ in this ordering, we have ${\effug_{\it{att}}(ug_j)}$ = $\it{att} (ug_j)$, giving us base cases for this recursive definition. For simplicity, we defined \effatt($ug$) = \effug$_{att}$($ug$) for $ug$ $\in$ \UG.  Therefore, for attribute roomAcc, effective values for user group $\mathrm{G_2}$ is $ e\mathrm{\_{roomAcc}(G_2)}$ = \{3.02\}. This value is same as its direct value for roomAcc attribute, since $\mathrm{G_2}$ has no junior group in UGH. For user group $\mathrm{G_1}$, $ e\mathrm{\_{roomAcc}(G_1)}$ = \{2.03, 2.04, 3.02\} as it inherits values from $\mathrm{G_2}$ and $\mathrm{G_3}$. The function \effu$_{att}$ maps the user to the effective values for attribute $att$, which is the union of its direct values and the effective values of $att$ for all its direct groups. For convenience we defined \effatt($u$) = \effu$_{att}$($u$) for user $u$ $\in$ \U and as shown in  Table \ref{tab:hgabac}, the effective values for attribute roomAcc for user Bob, written as $ e\mathrm{\_{roomAcc}(Bob)}$ = \{1.2, 2.03, 2.04, 3.02\} which is the union of its directly assigned value for roomAcc and values inherited from group $\mathrm{G_1}$. Similarly other effective attributes for user Bob can be calculated. The prime benefit of HGABAC model, which is easy assignment of multiple attributes to a user with user group memberships, is reflected in this function where user u is assigned multiple attributes with direct group membership of G$_1$.

Subject s $\in$ S created by the user u $\in$ U will then assume a subset of all effective attributes of user u. Similar effective attributes can be assigned to objects, which is out of scope of our reachability analysis and is not discussed. Authorization policies are pre-defined in the system, using propositional logic formula, for each operation in OP (set of operations) by security administrators, which determine if a subject is allowed to perform operations on objects, based on their effective attributes.

\begin{table}[t]
\centering
\caption{Example Configuration as Defined in Fig 2}
\label{tab:hgabac}
\resizebox{\columnwidth}{!}{%
\begin{tabular}{llllll}
\toprule
\multicolumn{6}{l}{\textbf{Basic Sets and Relations}}\\
\multicolumn{6}{l}{\begin{tabular}[c]{@{}l@{}}-- \U = \{Bob\}, \;\;UG =$\mathrm{ \{G_1, G_2, G_3\}}$.\end{tabular}} \\
\multicolumn{6}{l}{\begin{tabular}[c]{@{}l@{}}
-- \ATTR = \{skills, roomAcc, studType, college\}.\end{tabular}} \\

\multicolumn{6}{l}{-- SCOPE of each $\mathrm{att}$ in \ATTR, denoted by SCOPE$_{att}$: }\\
\multicolumn{6}{l}{\begin{tabular}[c]{@{}l@{}}\;\; studType = \{Grad, UnderGrad\},\qquad college = \{COS, COE, BUS\} \\\;\; skills = \{c, c++, java\},\qquad\qquad roomAcc = \{1.2, 2.03, 2.04, 3.02\}.\end{tabular}}\\
\multicolumn{6}{l}{-- UGH is given in Figure 2, highlighted in gray area.}\\

\multicolumn{6}{l}{\textbf{Direct Attributes} }\\

\multicolumn{6}{l}{\begin{tabular}[c]{@{}l@{}}\;\;skills(Bob) = \{c, java\}, \qquad\qquad roomAcc(Bob) = \{1.2\},\end{tabular}}\\

\multicolumn{6}{l}{\begin{tabular}[c]{@{}l@{}}\;\;roomAcc(G$_2$) = \{3.02\}, \qquad\qquad college(G$_2$) = \{COS\},\\\;\;roomAcc(G$_3$) = \{2.04\}, \qquad\qquad studType(G$_1$) = \{Grad\}, \\\;\;roomAcc(G$_1$) = \{2.03\}.\end{tabular}}\\

\multicolumn{6}{l}{\begin{tabular}[c]{@{}l@{}}\textbf{Direct User Groups}\;\; \\ \;\;\directug(Bob) = \{G$_1$\}. \end{tabular}}\\
\multicolumn{6}{l}{\begin{tabular}[c]{@{}l@{}}\textbf{Effective User Groups}\;\; \\ \;\;\effectiveug(Bob) = \{G$_1$, G$_2$, G$_3$\}\end{tabular}}\\

\multicolumn{6}{l}{\textbf{Effective User Group Attributes} }\\

\multicolumn{6}{l}{\begin{tabular}[c]{@{}l@{}}\;\;$e\mathrm{\_{roomAcc}(G_2)}$ = \{3.02\}, \qquad\qquad $ e\mathrm {\_{college}(G_2)}$ = \{COS\}, \\\;\;$ e\mathrm {\_{roomAcc}(G_3)}$ = \{2.04\}, \qquad\qquad $ e\mathrm {\_{studType}(G_1)}$ = \{Grad\}, \\\;\;$ e\mathrm{\_{roomAcc}(G_1)}$ = \{2.03, 2.04, 3.02\}, \qquad
$ e\mathrm{\_{college}(G_1)}$ = \{COS\}.\end{tabular}}\\

\multicolumn{6}{l}{\textbf{Effective User Attributes }}\\

\multicolumn{6}{l}{\begin{tabular}[c]{@{}l@{}}\;\;$ e\mathrm {\_{skills}(Bob)}$ = \{c,~java\}, \\\;\;$ e\mathrm {\_{roomAcc}(Bob)}$ = \{1.2,~2.03,~2.04,~3.02\}, \\\;\;$ e\mathrm {\_{studType}(Bob)}$ = \{Grad\}, \;\;$ e\mathrm {\_{college}(Bob)}$ = \{COS\}.\end{tabular}}\\

\bottomrule

\end{tabular}
}
\end{table} 

\textbf{Note:} HGABAC only allows set-valued attributes.  ABAC models generally allow set-valued as well as atomic-valued attributes (for example~\cite{jin2012unified}).  Inheritance of values via group membership for an atomic-valued attribute is problematic since such attributes can have only one value.  Hence, while the \GURA administrative model allows both atomic and set valued attributes the HGABAC only allows set values.

\subsection{GURA\textsubscript{G} Administrative Model}
The \GURAG administrative model \cite{gupta2016mathrm} was proposed to regulate the assignment of user attribute values in HGABAC model via direct user attributes, user-group attributes and user to group memberships. For convenience we understand the term ``assignment of attributes'' to mean ``assignment of attribute values.''  The model is inspired by ARBAC97 \cite{sandhu1999arbac97} and \GURA \cite{jin2012role} administrative models, where administrative roles and current attributes of user and groups or user to group memberships are considered to make future attributes or groups assignments. Administrative role hierarchy also exists in the system where senior administrator roles inherit permissions from junior roles.  The \GURAG model has three sub models (shown in Figure \ref{fig:hgabac}): user attribute assignment (UAA), user group attribute assignment (UGAA) and user to group assignment (UGA), which regulates the direct and effective attributes of users.  It should be noted that user group hierarchy (UGH) is considered fixed in the system and is not modified. Each of these sub models have different sets of administrative relations and preconditions definition using  policy language as discussed in following subsections.

The main difference between \GURA and \GURAG is that \GURAG includes the assignment of attributes to groups and user to group memberships. Further, the prerequisite conditions specified in \GURAG are more expressive, as it also checks the current effective attributes or effective group memberships of entities to make future assignments.

\begin{table}
  \centering
  \caption{Administrative Requests}\label{tab 2}
  \resizebox{\columnwidth}{!}{%
  \begin{tabular}{|ll|}
  \hline
\multicolumn{2}{|l|}{\begin{tabular}[c]{@{}l@{}}In the following requests:\\ $ar \in \mathrm{AR},\; att \in \mathrm{\ATTR},\; val \in \SCOPE_{att},\; u \in \U,\; ug \in \UG$\end{tabular}}                                                                                                                                                                                                                                                                                                                                                                                                                                                                             \\[1mm]
\toprule \hline
\multicolumn{2}{|c|}{\leftline{-- For User Attributes}}\\
 &\multicolumn{1}{c|}{\centerline{${\mathsf{add}(ar,~u,~att,~val)}$}}                                                                                           \\ 
& \multicolumn{1}{c|}{${\mathsf{delete}(ar,~u,~att,~val)}$}                                                                                           \\[1mm] 
\multicolumn{2}{|c|}{\leftline{-- For User Group Attributes}}\\
& \multicolumn{1}{c|}{${\mathsf{add}(ar,~ug,~att,~val)}$}                                                                                           \\ 
& \multicolumn{1}{c|}{${\mathsf{delete}(ar,~ug,~att,~val)}$}                                                                                           \\[1mm] 
\multicolumn{2}{|c|}{\leftline{-- For User to User-Group Membership }}\\
& \multicolumn{1}{c|}{${\mathsf{assign}(ar,~u,~ug)}$}                                                                                           \\ 
& \multicolumn{1}{c|}{${\mathsf{remove}(ar,~u,~ug)}$}                                                                                           \\ 
\bottomrule
  \end{tabular}
  }
\end{table} 
\subsubsection{Administrative Requests}

\begin{mydef} [\textbf{Administrative Requests}] \label{def:adminrequest}
The attributes and group memberships of entities are changed by administrative request made by administrators with certain administrative roles as defined in {Table \ref{tab 2}}, where AR is the finite set of administrative roles. The administrative request ${\mathsf{add}(ar,~u,~att,~val)}$ is made by administrator with role $ar$ to add value $val$ to attribute $att$ of user $u$. Similar administrative request are used for groups also. Administrative requests $\mathsf{assign}$ and $\mathsf{remove}$ are required for managing group memberships. Each administrative request can add or delete a single attribute value from a user or group.
\end{mydef}

\subsubsection{Administrative Rules}
\begin{table}[t]
\centering
\caption{\GURAG Administrative Model }
\label{tab:gurag-model1}
{%
\begin{tabularx}\columnwidth{XXXXXX}
\toprule
\multicolumn{6}{l}{\begin{tabular}[c]{@{}l@{}}-- User Attribute Assignment (\textbf{UAA}):\\\;\; For each $att$ in UA,\end{tabular}}\\[\defaultaddspace]

\multicolumn{6}{l}{\begin{tabular}[c]{@{}l@{}}\centerline{$\canA_{att} \subseteq \AR \times \C \times \SCOPE_{att}$}\end{tabular}}\\[1mm]
\multicolumn{6}{l}{\begin{tabular}[c]{@{}l@{}}\centerline{$\canD_{att} \subseteq \AR \times \C \times \SCOPE_{att}$}\end{tabular}}\\[1mm]
\multicolumn{6}{l}{\begin{tabular}[c]{@{}l@{}}-- User Group Attribute Assignment (\textbf{UGAA}):\\\;\; For each $att$ in UA,\end{tabular}}\\[\defaultaddspace]

\multicolumn{6}{l}{\begin{tabular}[c]{@{}l@{}}\centerline{$\canAUG_{att} \subseteq \AR \times \C \times \SCOPE_{att}$}\end{tabular}}\\[1mm]
\multicolumn{6}{l}{\begin{tabular}[c]{@{}l@{}}\centerline{$\canDUG_{att} \subseteq \AR \times \C \times \SCOPE_{att}$}\end{tabular}}\\[1mm]

\multicolumn{6}{l}{-- User to User Group Assignment (\textbf{UGA}):}\\[\defaultaddspace]

\multicolumn{6}{l}{\begin{tabular}[c]{@{}l@{}}\centerline{$\Assign  \subseteq \AR \times \C \times \UG$}\end{tabular}}\\[1mm]
\multicolumn{6}{l}{\begin{tabular}[c]{@{}l@{}}\centerline{$\Remove  \subseteq \AR \times \C \times \UG$}\end{tabular}}\\[1mm]

\bottomrule

\end{tabularx}
}
\end{table} 
\begin{mydef}[\textbf{Administrative Rules}] \label{def:adminrule}
Administrative rules are tuples in administrative relations which specify conditions under which administrative requests are authorized. Each of the three sub-models (UAA, UGAA, UGA) in \GURAG model have administrative relations to define these rules.

The UAA sub-model deals with addition or deletion of attributes from the user. It has two administrative relations shown in Table \ref{tab:gurag-model1}, where a rule $\langle ar,~c,~val\rangle$ $\in$ $\canA_{att}$ authorizes request $\mathsf{add}(ar$,$~u$,$~att$,$~val)$ if user $u$ satisfies precondition $c$. Similarly, rule $\langle ar,~c,~val\rangle$ $\in$ $\canD_{att}$ authorizes $\mathsf{delete}(ar$,$~u$,$~att$,$~val)$ requests if user $u$ satisfies precondition $c$. In UAA, the precondition $c$ $\in$ \C~includes only current direct and effective attributes of user $u$. Similar relations also exist for administering attributes of user groups as discussed in sub-model UGAA. In UGAA, $c$ $\in$ \C~involves current direct or effective attributes of the group whose attributes are modified.

The UGA sub-model has two relations shown in lower part of Table \ref{tab:gurag-model1}. The rule $\langle ar,~c,~ug\rangle$ $\in$ $\Assign$ authorizes user to group assignment request $\mathsf{assign}(ar$,$~u$,$~ug$) if user $u$ satisfies the precondition $c$. Similarly rule $\langle ar,~c,~ug\rangle$ $\in$ $\Remove$ authorizes remove request $\mathsf{remove}(ar$,$~u$,$~ug)$ if user $u$ satisfies precondition $c$. The precondition $c$ $\in$ \C~involves both current direct or effective attributes and groups of user $u$.

The expressive power of the \GURAG model is primarily determined by the richness of the policy language used to define the preconditions C in Table \ref{tab:gurag-model1}.  The most general language for this purpose is defined in \cite{gupta2016mathrm}, similar to the most general language of \cite{jin2012role} (but without atomic attributes).

\textbf{Note:} In the original \GURAG definition~\cite{gupta2016mathrm}, the administrative relations of Table~\ref{tab:gurag-model1} are defined with $2^{\SCOPE_{att}}$ substituted for $\SCOPE_{att}$ and $2^{\UG}$ substituted for $\UG$.  With the modification of Table~\ref{tab:gurag-model1} the administrative relations can grow linearly in the size of $\SCOPE_{att}$ and $\UG$.  This does not materially impact the complexity analysis of the reachability problem.
\end{mydef}

\subsubsection{GURA\textsubscript{G} scheme}
For purpose of our reachability analysis, we express the \GURAG model according to the notations developed in \cite{tripunitara2007theory}, following the treatment in~\cite{jin2013reachability}. The \GURAG scheme is presented as a state transition system where each state consists of direct attribute assignments for each attribute of every user and group, and also each user to groups membership. A transition between states occurs when an authorized administrative request changes either direct user or group attribute, or changes user to group membership. The general definition for \GURAG scheme is as follows.

\begin{mydef}[\GURAG\xspace \textbf{Scheme}] \label{def:scheme}
A \GURAG\xspace scheme is a state transition system $\langle$$\U$, $\ATTR$, $\AR$, $\SCOPE$, $\UG$, $\succeq_{ug}$, $\Psi$, $\Gamma$, $\delta$$\rangle$ where,

\begin{enumerate}[label=(\roman*)]
\item \U, \ATTR, \AR, \UG, $\succeq_{ug}$  are as defined in Tables \ref{tab:hgabac-model} and \ref{tab 2}.
\item $\SCOPE\xspace= \langle \SCOPE_{att_1}\ldots\SCOPE_{att_n}\rangle$ where $att_i$ $\in$ \ATTR, is the collection of scopes of all attributes.
\item $\Psi$ is the collection of all administrative rules in UAA, UGAA and UGA sub-models.
\item $\Gamma$ and $\delta$ are set of states and transition function respectively, defined in following parts of this subsection.
\end{enumerate}
\begin{table*}[t]
\centering
\caption{Transition Function}
\label{tab transition}
\renewcommand{\arraystretch}{1.1}

\begin{tabular}{l|l|l}
\hline
\multicolumn{3}{|p{.99\textwidth}|}{(1) $\gamma_1$ and $\gamma_2$ are the source and target states respectively.}                                                                                                                                                                                                                                                                                                                                                                                                                                                                             \\
\multicolumn{3}{|p{.99\textwidth}|}{(2) Let : $ar$ $\in$ \AR,\; $u$ $\in$ \U,\; $ug$ $\in$ \UG,\; $att$ $\in$ \ATTR,\; $val'$ $\in$ ${\SCOPE_{att}}$,\; $ug' \in \UG$.}\\
\multicolumn{3}{|p{.99\textwidth}|}{
(3)
$\Evaluate_u$: \U $\times$ \C $\times$ $\Gamma$
$\rightarrow$ \{\true, \false\}, returns \true\xspace
if user $u\in \U$ satisfies precondition $c\in \C$ in state $\gamma\in\Gamma$, else \false.}
\\
\multicolumn{3}{|p{.99\textwidth}|}{\begin{tabular}[c]{@{}l@{}}
(4)
$\Evaluate_{ug}$ : \UG $\times$ \C $\times$ $\Gamma$
$\rightarrow$ \{\true, \false\}, returns \true\xspace
if user group $ug\in \UG$ satisfies precondition $c\in \C$ in state $\gamma\in\Gamma$,\\ $\hspace{165 mm}$else \false.\end{tabular}
}
\\
\multicolumn{3}{|p{.99\textwidth}|}{
(5)
$\Evaluate_{u\?{12}-\?{12}ug}$ : \U $\times$ \C $\times$ $\Gamma$
$\rightarrow$ \{\true, \false\}, returns \true\xspace
if user $u\in \U$ satisfies precondition $c\in \C$ in state $\gamma\in\Gamma$, else \false .
}\\
\hline
\hline
\multicolumn{1}{|c|}{\textbf{Request}}                                                                                                           & \multicolumn{1}{c|}{\textbf{Pre-Conditions}}                                                                                           & \multicolumn{1}{c|}{\textbf{Target State}}                                                                                                                                                                                                                                                                                                         \\ \hline
\begin{tabular}[c]{@{}l@{}}$\mathsf{add}$($ar$, $u$, $att$, $val^\prime$) \end{tabular}   & \begin{tabular}[c]{@{}l@{}}
$\exists$ $\mathrm{\langle}ar,~c,~val'\mathrm{\rangle}$ $\in$ $\mathsf{canAddU}_{att}$. \\
( $\Evaluate_u(u,~c,~\gamma_1) \; \land$ \\ $  val^\prime \notin att_{\gamma_1}(\mathit{u})$)\end{tabular}& \begin{tabular}[c]{@{}l@{}}$att_{\gamma_2}(\mathit{u})$ = $ att_{\gamma_1}(\mathit{u}) \;\!\cup\! \;\{val'\},$ \\
$att_{\gamma_2}(\mathit{ug})$ = $ att_{\gamma_1}(\mathit{ug}),$
$\directug_{\gamma_2}(u)$ = ${\directug_{\gamma_1}(u)}$,\\
$\UAAN_{\gamma_2}$ = $\UAAN_{\gamma_1}$ $\backslash$ $\mathrm{\langle}$$u$, $att$,
$att_{\gamma_{1}}(u)$$\mathrm{\rangle}$ $\cup$ $\mathrm{\langle}$$u$, $att$,
$att_{\gamma_{2}}(u)$$\mathrm{\rangle}$.
\end{tabular}                                                                                                                                                                                                                                                                                 \\  \hline

\begin{tabular}[c]{@{}l@{}}$\mathsf{delete}$($ar$, $u$, $att$, $val^\prime$)\end{tabular}   & \begin{tabular}[c]{@{}l@{}}
$\exists$ $\mathrm{\langle}ar,~c,~val'\mathrm{\rangle}$ $\in$ $\mathsf{canDeleteU}_{att}$. \\
( $\Evaluate_u(u,~c,~\gamma_1) \; \land $ \\ $  val^\prime \in att_{\gamma_1}(\mathit{u})$)\end{tabular}& \begin{tabular}[c]{@{}l@{}}$att_{\gamma_2}(\mathit{u})$ = $ att_{\gamma_1}(\mathit{u}) \;\!\setminus\!\;\{val'\},$\\
$att_{\gamma_2}(\mathit{ug})$ = $ att_{\gamma_1}(\mathit{ug}),$
$\directug_{\gamma_2}(u)$ = ${\directug_{\gamma_1}(u)}$,\\
$\UAAN_{\gamma_2}$ = $\UAAN_{\gamma_1}$ $\backslash$ $\mathrm{\langle}$$u$, $att$,
$att_{\gamma_{1}}(u)$$\mathrm{\rangle}$ $\cup$ $\mathrm{\langle}$$u$, $att$,
$att_{\gamma_{2}}(u)$$\mathrm{\rangle}$.
\end{tabular}                                                                                                                                                                                                                                                                                 \\  \hline

\begin{tabular}[c]{@{}l@{}}$\mathsf{add}$($ar$, $ug$, $att$, $val^\prime$)\end{tabular}   & \begin{tabular}[c]{@{}l@{}} $\exists$ $\mathrm{\langle}ar,~c,~val'\mathrm{\rangle}$ $\in$ $\mathsf{canAddUG}_{att}$. \\
( $\Evaluate_{ug}(ug,~c,~\gamma_1) \; \land$ \\ $  val^\prime \notin att_{\gamma_1}(\mathit{ug}) $) \end{tabular}& \begin{tabular}[c]{@{}l@{}}$att_{\gamma_2}(\mathit{ug})$ =$ att_{\gamma_1}(\mathit{ug})\;\!\cup\!\;\{val'\},$ \\
$att_{\gamma_2}(\mathit{u})$ = $ att_{\gamma_1}(\mathit{u}),$
$\directug_{\gamma_2}(u)$ = ${\directug_{\gamma_1}(u)}$,\\
$\UGAAN_{\gamma_2}$ = $\UGAAN_{\gamma_1}$ $\backslash$ $\mathrm{\langle}$$ug$, $att$,
$att_{\gamma_{1}}(ug)$$\mathrm{\rangle}$ $\cup$ $\mathrm{\langle}$$ug$, $att$,
$att_{\gamma_{2}}(ug)$$\mathrm{\rangle}$.
\end{tabular}                                                                                                                                                                                                                                                                                 \\  \hline
\begin{tabular}[c]{@{}l@{}}$\mathsf{delete}$($ar$, $ug$, $att$, $val^\prime$)\end{tabular}   & \begin{tabular}[c]{@{}l@{}} $\exists$ $\mathrm{\langle}ar,~c,~val'\mathrm{\rangle}$ $\in$ $\mathsf{canDeleteUG}_{att}$. \\
( $\Evaluate_{ug}(ug,~c,~\gamma_1) \; \land$ \\ $ val^\prime \in att_{\gamma_1}(\mathit{ug}) $)  \end{tabular}& \begin{tabular}[c]{@{}l@{}}$att_{\gamma_2}(\mathit{ug})$ = $ att_{\gamma_1}(\mathit{ug})\;\!\setminus\!\;\{val'\},$
\\
$att_{\gamma_2}(\mathit{u})$ = $ att_{\gamma_1}(\mathit{u}),$
$\directug_{\gamma_2}(u)$ = ${\directug_{\gamma_1}(u)}$,\\
$\UGAAN_{\gamma_2}$ = $\UGAAN_{\gamma_1}$ $\backslash$ $\mathrm{\langle}$$ug$, $att$,
$att_{\gamma_{1}}(ug)$$\mathrm{\rangle}$ $\cup$ $\mathrm{\langle}$$ug$, $att$,
$att_{\gamma_{2}}(ug)$$\mathrm{\rangle}$.

\end{tabular}                                                                                                                                                                                                                                                                                 \\
\hline

\begin{tabular}[c]{@{}l@{}}$\mathsf{assign}$($ar$, $u$, $ug'$)\end{tabular}   & \begin{tabular}[c]{@{}l@{}}$\exists$ $\mathrm{\langle}ar,~c,~ug'\mathrm{\rangle}$ $\in$ $\mathsf{canAssign}$.\\
( $\Evaluate_{u\?{12}-\?{12}ug}(u,~c,~\gamma_1) \; \land$ \\ $ ug^\prime \notin \directug_{\gamma_1}(u) $) \end{tabular}& \begin{tabular}[c]{@{}l@{}}$\directug_{\gamma_2}(u)$ = ${\directug_{\gamma_1}(u)\;\!\cup\!\;\{ug'}\}$  \\
$att_{\gamma_2}(\mathit{u})$ = $ att_{\gamma_1}(\mathit{u}),$
$att_{\gamma_2}(\mathit{ug})$ = $ att_{\gamma_1}(\mathit{ug}),$\\
$\UGAN_{\gamma_2}$ = $\UGAN_{\gamma_1}$ $\backslash$ $\mathrm{\langle}$$u$,
$\directug_{\gamma_1}(u)$$\mathrm{\rangle}$ $\cup$ $\mathrm{\langle}$$u$,
$\directug_{\gamma_2}(u)$$\mathrm{\rangle}$.
\end{tabular}                                                                                                                                                                                                                                                                                 \\  \hline

\begin{tabular}[c]{@{}l@{}}$\mathsf{remove}$($ar$, $u$, $ug'$)\end{tabular}   & \begin{tabular}[c]{@{}l@{}}$\exists$ $\mathrm{\langle}ar,~c,~ug'\mathrm{\rangle}$ $\in$ $\mathsf{canRemove}$.\\
( $\Evaluate_{u\?{12}-\?{12}ug}(u,~c,~\gamma_1) \; \land$ \\ $ ug^\prime \in \directug_{\gamma_1}(u)$)\end{tabular}& \begin{tabular}[c]{@{}l@{}}$\directug_{\gamma_2}(u)$ = ${\directug_{\gamma_1}(u)\;\!\setminus\!\;\{ug'}\}$\\
$att_{\gamma_2}(\mathit{u})$ = $ att_{\gamma_1}(\mathit{u}),$
$att_{\gamma_2}(\mathit{ug})$ = $ att_{\gamma_1}(\mathit{ug}),$ \\
$\UGAN_{\gamma_2}$ = $\UGAN_{\gamma_1}$ $\backslash$ $\mathrm{\langle}$$u$,
 $\directug_{\gamma_1}(u)$$\mathrm{\rangle}$ $\cup$ $\mathrm{\langle}$$u$,
$\directug_{\gamma_2}(u)$$\mathrm{\rangle}$.
 \end{tabular}                                                                                                                                                                                                                                                                                 \\  \hline

\end{tabular}
\end{table*} 
\subsubsection{Direct State}
$\Gamma$ is the finite set of states where each state $\gamma$ $\in$ $\Gamma$ records directly assigned attributes of each user and user group, along with user to groups membership. The direct user attribute assignment in state $\gamma$, denoted by \UAAN$_\gamma$, contains tuples of the form $\langle u, att, val\rangle$ for every $u$ $\in$ \U and every $att$ $\in$ \ATTR such that $att(u)=val$ and $val \in $ \Range($att$) in state $\gamma$.  To ensure uniqueness of user attribute values we require the following.
\begin{flalign*}
\langle u, att, val_1\mathrm\rangle \in \UAAN_\gamma \wedge \langle u, att, val_2\mathrm\rangle \in \UAAN_\gamma \Rightarrow val_1=val_2
\end{flalign*}
\noindent Similarly, direct user group attribute assignment in state $\gamma$, denoted by \UGAAN$_\gamma$, contains tuples of the form $\langle ug, att, val\rangle$ for every $ug$ $\in$ \UG and every $att$ $\in$ \ATTR such that $att(ug)=val$ and $val \in \Range(att)$ in state $\gamma$, with the following uniqueness requirement.
\begin{flalign*}
\langle ug, att, val_1\mathrm\rangle \in \UGAAN_\gamma \wedge \langle ug, att, val_2\mathrm\rangle \in \UGAAN_\gamma \Rightarrow val_1=val_2
\end{flalign*}
\noindent Finally, direct user to group assignment in state $\gamma$, denoted \UGAN$_\gamma$, contains tuples of the form $\langle u, val\rangle$ for every $u$ $\in$ \U such that $\directug(u)=val$ and $val\in 2^\UG$ in state $\gamma$, with the following uniqueness requirement.
\begin{flalign*}
\langle u, val_1\mathrm\rangle \in \UGAN_\gamma \wedge \langle ug, val_2\mathrm\rangle \in \UGAN_\gamma \Rightarrow val_1=val_2
\end{flalign*}

Note that information in a state can be used to calculate the effective attributes for user or group and effective user to groups membership in that state.  For convenience we understand the notation $att_\gamma(u)$, $att_\gamma(ug)$ and $\directug_\gamma(u)$ to denote the values of these functions in state $\gamma$ for $u\in$ U and $ug\in$ UG.

\subsubsection{Transition Function}
Any change in the direct state records $\mathrm(\UAAN_\gamma, \UGAAN_\gamma, \UGAN_\gamma)$ will transform the current state to a new state.  The transition function specifies the change from one state to another in a \GURAG system based on current direct or effective values and administrative requests, as shown in Table \ref{tab transition}. Formally,
$\delta: \Gamma \times \REQ \rightarrow \Gamma$, where \REQ is the set of possible administrative requests.
\end{mydef} 

\section{Restricted GURA\textsubscript{G} (\upshape{r}GURA\textsubscript{G})}
\label{s:restricted}
In this section, we introduce a restricted form of \GURAG administrative model, called \miniGURANG, used in our attribute reachability analysis. This restricted form allows a subset of the precondition language defined for \GURAG \cite{gupta2016mathrm}, whereby our analysis also establishes lower bound results on the complexity analysis for richer \GURAG model.  We first present a generalized policy language for \miniGURANG, followed by three specific instances---\miniGURANGZ, \miniGURANGO, and \miniGURANGT. 

The left side of Figure \ref{fig:rgurag} shows the relation between these schemes, while the right side shows the \miniGURA~schemes discussed in \cite{jin2013reachability}. At a high level, \miniGURANGZ and \miniGURANGO add group attributes respectively to \miniGURANZ and \miniGURANO, while \miniGURANGT further adds administration of user membership in groups.
Thereby, in \miniGURANGZ and \miniGURANGO the administrative relations $\Assign$ and $\Remove$ are empty whereas they are populated in \miniGURANGT.
Table \ref{tab ex-rgurag} provides example administrative rules for each \rGURAG instance, as will be explained below.

\begin{mydef}[\rGURAG \textbf{Scheme}]
The \rGURAG scheme uses the policy grammar below, to specify preconditions C in Table \ref{tab:gurag-model1},
\begin{itemize}
 \item[] $\varphi$ ::=$~\neg~\varphi~|~\varphi\wedge\varphi~|~svalue\in$ $\mathrm{direct}~| ~svalue\in \mathrm{effective}$
 \item[] $svalue ::=~sval_1~|~sval_2~|~\ldots~|~sval_m$
\end{itemize}
where \SCOPE$_{att}=\{sval_1,~sval_2,~\ldots,~sval_m\}$.  The two non-terminals $\mathrm{direct}$ and $\mathrm{effective}$, are individually defined in its three instances---\miniGURANGZ, \miniGURANGO and \miniGURANGT---in following subsections.


\begin{table*}[t]
\centering
\caption{Example Rules in \miniGURANGZ, \miniGURANGO and \miniGURANGT Schemes }
\label{tab ex-rgurag}
\begin{tabular}{|c|c|c|c|}
\cline{1-4}
{\textbf{Relation}} & {\textbf{Admin Role}} &  {\textbf{Pre-requisite Condition}} & {\textbf{Value}} \\ \hline
\multicolumn{4}{|c|}{Rules in \miniGURANGZB scheme }                                                                                                                                                                                                                                                                                                                                                                                                                                                                            \\ \cline{1-4}

$\canA_{skills}$ &DeptAdmin                                 & \begin{tabular}[c]{@{}c@{}} c $\in  {e\_{{skills}}(u)}$ \;$\land $ $\neg$ (java $\in$ ${{\textit{skills}}(u)}$)\end{tabular}                                          & c++                                       \\ 
$\canD_{roomAcc}$ &BuildAdmin                                 & \begin{tabular}[c]{@{}c@{}} 3.02 $\in  {e\_{{roomAcc}}(u)}$ \end{tabular}                                          & 1.2                                       \\ \cline{1-4}
$\canAUG_{college}$ &UnivAdmin                                 & \begin{tabular}[c]{@{}c@{}} $\neg$ (COE $\in$ ${{\textit{college}}(ug)}$)\end{tabular}                                          & COS                                       \\ 
$\canDUG_{roomAcc}$ &BuildAdmin                                 & \begin{tabular}[c]{@{}c@{}} 2.04 $\in  {e\_{{roomAcc}}(ug)}$ \end{tabular}                                          & 2.03                                       \\ \hline

\multicolumn{4}{|c|}{Rules in \miniGURANGOB scheme further add}                                                                                                                                                                                                                                                                                                                                                                                                                                                                            \\
\cline{1-4}

$\canA_{studType}$ &DeptAdmin                                 & \begin{tabular}[c]{@{}c@{}} java $\in  {e\_{{skills}}(u)}$ \;$\land $ 2.03 $\in$ ${{\textit{roomAcc}}(u)}$\end{tabular}                                          & Grad                                       \\ 
$\canD_{roomAcc}$ &BuildAdmin                                 & \begin{tabular}[c]{@{}c@{}} 3.02 $\in  {{{roomAcc}}(u)}$ \;$\land $ COS $\in$ ${{\textit{college}}(u)}$ \end{tabular}                                          & 3.02                                       \\ \cline{1-4}
$\canAUG_{skills}$ &DeptAdmin                                 & \begin{tabular}[c]{@{}c@{}}  COS $\in$ ${{\textit{college}}(ug)}$ $\land $ UnderGrad  $\in$ ${{e\_{studType}}(ug)}$\end{tabular}                                          & java                                       \\ 
$\canDUG_{college}$ &UnivAdmin                                 & \begin{tabular}[c]{@{}c@{}} 2.04 $\in  {e\_{{roomAcc}}(ug)}$ $\land $ 2.03 $\in  {e\_{{roomAcc}}(ug)}$ \end{tabular}                                          & BUS                                       \\ \hline

\multicolumn{4}{|c|}{Rules in \miniGURANGTB scheme further add}                                                                                                                                                                                                                                                                                                                                                                                                                                                                            \\
\cline{1-4}
$\Assign$ &DeptAdmin                                 & \begin{tabular}[c]{@{}c@{}} 1.02 $\in  {e\_{{roomAcc}}(u)}$ $\land $  $\neg$ (BUS $\in$ ${{\textit{college}}(u)}$) $\land $  $\mathrm{G_2}$ $\in$  \directug($u$)  \end{tabular}                                          & $\mathrm{G_1}$                                      \\ 
$\Remove$ &GroupAdmin                                 & \begin{tabular}[c]{@{}c@{}} $\mathrm{G_1}$ $\in$ \effectiveug($u$) $\land$ $\mathrm{G_2}$ $\in$ \directug($u$)  \end{tabular}                                          & $\mathrm{G_2}$                                       \\ \hline

\end{tabular}
\end{table*}

\subsection{The \upshape{r}GURA\textsubscript{G\textsubscript{0}} Scheme}\label{subsec:rgura0}
In \miniGURANGZ scheme, preconditions for rules in $\mathsf{canAddU}_{att}$ and $\mathsf{canDeleteU}_{att}$ relations only allow the same attribute $att$ whose value is added or deleted from user. Therefore, conditions for user $u$ have non-terminals $\mathrm{direct}$ and $\mathrm{effective}$ defined as follows.
\begin{itemize}
  \item[] $\mathrm{direct::}$= $att(u)$ \;\;\;\;\&\;\;\;\;    $\mathrm{effective::}$= $\effatt(u)$
\end{itemize}
Similarly, the administrative relations in $\mathsf{canAddUG}_{att}$ and $\mathsf{canDeleteUG}_{att}$ for user group $ug$ have $\mathrm{direct}$ and $\mathrm{effective}$ defined as follows.
\begin{itemize}
  \item[] $\mathrm{direct::}$= $att(ug)$   \;\;\;\;\&\;\;\;\;    $\mathrm{effective::}$= $\effatt(ug)$
\end{itemize}
The examples for \miniGURANGZ shown in Table \ref{tab ex-rgurag} conform to these restrictions.  Note that the attribute being updated is given as the subscript in the Relation column and the conditions in the Pre-requisite Condition column only involve this attribute.

\subsection{The \upshape{r}GURA\textsubscript{G\textsubscript{1}} Scheme}\label{subsec:rgura1}
In \miniGURANGO scheme, the precondition can include any attribute from the set of attributes.
Therefore, conditions in rules for $\mathsf{canAddU}_{att}$ and $\mathsf{canDeleteU}_{att}$ relations for user $u$ have $\mathrm{direct}$ and $\mathrm{effective}$ defined as follows where $att_i$ $\in$ \ATTR.
\begin{itemize}
  \item[] $\mathrm{direct::}$= $att_i(u)$ \;\;\;\&\;\;\;\;\;    $\mathrm{effective::}$= $\effatt_i(u)$
\end{itemize}
Similarly, the conditions for user group $ug$ in relations $\mathsf{canAddUG}_{att}$ and $\mathsf{canDeleteUG}_{att}$ have non-terminals $\mathrm{direct}$ and $\mathrm{effective}$ defined as follows.
\begin{itemize}
  \item[] $\mathrm{direct::}$= $att_i(ug)$    \;\;\;\;\&\;\;\;\;    $\mathrm{effective::}$= $\effatt_i(ug)$
\end{itemize}
The added rules for \miniGURANO in Table \ref{tab ex-rgurag} illustrate this, where the preconditions involve attributes other than the one being updated. The earlier rules for \miniGURANGZ continue to be valid for \miniGURANO. 

\begin{figure}[t]
\centering
\includegraphics[scale=.60]{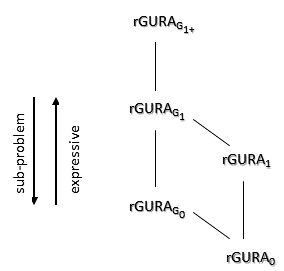}
\caption{\rGURAG (Left Side) and \rGURA (Right Side) Schemes}
\label{fig:rgurag}
\end{figure} 
\subsection{The \upshape{r}GURA\textsubscript{G\textsubscript{1+}} Scheme}\label{subsec:rgura1+}
The \miniGURANGT scheme allows changes in user group memberships besides modifying the attributes of user and user groups. Therefore, in addition to the grammar supported by \miniGURANGO scheme, \miniGURANGT also includes user's direct or effective group memberships as preconditions in rules for $\mathsf{canAssign}$ and $\mathsf{canRemove}$ administrative relations. The additional grammar to specify such preconditions is specified below:
\begin{itemize}
      \item[]  $\varphi ::= ug \in \directug(u)~ |~ug \in \effectiveug(u)$
     \end{itemize}
 \end{mydef}
\noindent
In Table \ref{tab ex-rgurag}, rule in $\mathsf{canAssign}$ includes effective values for $roomAcc$, direct values for $college$ attribute and direct groups of user $u$. 


\section{Reachability Problem Definition}
\label{s:problem}
In this section, we provide a formal definition of our attribute reachability problem along with the reachability query and different query types supported in our analysis.  The general approach is similar to that of \cite{jin2013reachability}, except that atomic-valued attributes are excluded (as noted in Section \ref{ss:hgabac}) and reachability is defined with respect to effective rather than direct attributes (\cite{jin2013reachability} does not have the notion of effective attributes).

The user attribute reachability analysis problem (or reachability problem) is based on the effective attributes of the user. Informally, the problem can be stated as: Given an initial transition system state with a set of attribute assignments of the user, the user's group memberships and the attributes of all the user's member groups, can administrators with a given set of administrative roles issue one or more administrative requests, which transition to a target state having the set of specified effective attributes for that user? We highlight some simplifications in our reachability analysis process. First, as the changes made to the attributes or group memberships of one user do not affect the attributes or group memberships of another user, our analysis will only determine the effective attributes of a single user of interest and hence will only consider attribute assignment of that user, its group memberships and attributes of these groups. Formally, we assume \U = \{u\} in our analysis \cite{jin2013reachability}.
Second, as the reachability analysis focuses on powers of a certain set of administrative roles \A\xspace$\subseteq$\xspace\AR, we do not consider the administrative rules specified for roles outside of \SUBAR. In other words, we can assume \AR = \SUBAR. These simplifications gives our analysis more convenient statements without loosing generality.

\begin{mydef} [\textbf{Reachability Query}] \label{def:query}
A reachability query $q$ $\in$ \Q~specifies a subset of effective values of a user for some attributes in any target state.
Formally,
\begin{itemize}[leftmargin=0mm]
\item[] $q$ $\subseteq$ \{$\mathrm{\langle}$$u$, $e\_{{att}}$, $vset$$\mathrm{\rangle}$ $|$ $u\in$ \U, $att\in$ \ATTR, $vset \in$ \Range($att$)\}
\end{itemize}
\end{mydef}
\noindent
In the example problems discussed in Section \ref{sec:example}, we will use the following notation to specify our query, which is equivalent to the notation defined above:
\begin{itemize}[leftmargin=0mm]
  \item [] $q$ $\subseteq$ \{${e\_att(u) = vset}$ $|$ $u\in$ \U, $att\in$ \ATTR, $vset \in$ \Range($att$)\}
\end{itemize}
\noindent
For example,\\ $q$ = $\mathrm{\{\langle u, e\_roomAcc,~ \{2.04\} \rangle, \langle u,  e\_skills,~ \{c\} \rangle}$ , \\$\mathrm{\;\;\;\;\;\;\;\; \langle u, e\_college,~ \{COS, COE\} \rangle\}}$   \qquad is equivalent to \\ $q$ = \{$\mathrm{e\_roomAcc(u) = \{2.04\}, e\_skills(u) = \{c\},}$ \\ $\mathrm{\;\;\;\;\;\;\;e\_college(u) = \{COS, COE\}}$\}.

Two types of reachability query are defined in the system. A query is called ``strict'' satisfied if every effective attribute value specified in the query is exactly the same as that in the target state. A query is called ``relaxed'' satisfied by the user if in the target state every effective attribute value of the user is a superset of the corresponding attribute values specified in the reachability query.
For example, let \ATTR = $\{$skills$\}$, \U = \{Bob\} and reachability query $q$ = $\langle$Bob, $e\_{skills}$, \{c, java\}$\rangle$. For strict query type, $q$ can be satisfied in states $\gamma^\prime  \in \Gamma$ where $e\_{skills}_{\gamma^\prime }(u)$ = \{c, java\}. In relaxed query type, $q$ can be satisfied by any state $\gamma''  \in \Gamma$ where $e\_{skills}_{\gamma''}(u)$ = $setval$ and $\mathrm{\{}$c, java$\mathrm{\}}$ $\subseteq$ $setval$. For ease of understanding, we represent the effective value of attribute $att$ for user $u$ in state $\gamma \in \Gamma$ as $e\_att_{\gamma}(u)$.
The formal definition for reachability query types is given below:
\begin{mydef} [\textbf{Reachability Query Types}]
 For any \rGURAG scheme $\langle$\U, \ATTR, \AR, \SCOPE, $\UG$, $\succeq_{ug}$, $\Psi$, $\Gamma$, $\delta$$\rangle$, we formally define two Reachability Query Types as:
\begin{itemize}
\item \RPSAME or strict satisfied queries have the entailment function $\vdash_{\RPSAME}$$\mathrm{:}$ $\Gamma$ $\times$ $\Q$ $\rightarrow$ $\mathrm{\{}$\true,
\false$\mathrm{\}}$ which returns \true\xspace (i.e., $\gamma$ $\vdash_{\RPSAME}$ $q$)
if
$\forall$ 
$\mathrm{\langle}$$u$, $e\_{{att}}$, $vset$$\mathrm{\rangle}$
$\in$ $q$. $\effatt_{\gamma}(u) =$ $vset$. 
\item \RPSUPER or relaxed satisfied queries have the entailment function $\vdash_{\RPSUPER}$$\mathrm{:}$ $\Gamma$ $\times$ $\Q$ $\rightarrow$ $\mathrm{\{}$\true,
\false$\mathrm{\}}$
which returns \true\xspace (i.e., $\gamma$ $\vdash_{\RPSUPER}$ $q$) if
$\forall$ $\mathrm{\langle}$$u$, $e\_{{att}}$, $vset$$\mathrm{\rangle}$
$\in$ $q$. $\effatt_{\gamma}(u)$ $\supseteq$ $vset$.
\end{itemize}
\end{mydef}
\noindent
It is clear that given a scheme and problem instance, if \RPSAME query problem is satisfied then \RPSUPER problem is also satisfied, but not vice versa. The following two definitions are same as defined in \cite{jin2013reachability}, but we will state them for the sake of completeness.

\begin{mydef}[\textbf{Reachability Plan}] \label{def:plan}
 A Reachability Plan or plan is a sequence of authorized administrative requests to transition from initial state to the target state. For any \rGURAG scheme $\langle$\U, \ATTR, \AR, \SCOPE, $\UG$, $\succeq_{ug}$, $\Psi$, $\Gamma$, $\delta$$\rangle$ and states $\gamma_0$, $\gamma^\prime$ $\in$ $\Gamma$, reachability plan is a sequence of authorized requests
$\langle$ $req_1$, $req_2$, $\ldots$, $req_n$$\rangle$ where $req$$_i$ $\in$ \REQ $\mathrm{(}$1 $\leq$ $i$ $\leq$ $n$$\mathrm{)}$, to transition from an initial state $\gamma_0$ to target state $\gamma^\prime$ if:
$ \gamma_0 \overset{req_1}{\rightarrow} \gamma_1\overset{req_2}{\rightarrow}\gamma_2\ldots\overset{req_n}{\rightarrow} \gamma^\prime $.
The arrow denotes a successful transition from one state to another due to an administrative request $req_i$ authorized by rules in $\Psi$. We write $\gamma_0$ $\overset{plan_\Psi}{\rightsquigarrow}$ $\gamma^\prime$ to abbreviate the complete plan.
\end{mydef}

Informally, a reachability problem deals if there exists a reachability plan to transition from an initial state to some target state where the effective attribute values of the user satisfy a particular reachability query. Formally,

\begin{mydef}[\textbf{Reachability Problems}] \label{def:problem}
Given any \rGURAG scheme $\langle$\U, \ATTR, \AR, \SCOPE, $\UG$, $\succeq_{ug}$, $\Psi$, $\Gamma$, $\delta$$\rangle$, the attribute reachability problem is as follows: 
\begin{itemize}
\item \RPSAME or strict reachability problem instance \I is of the form $\langle$$\gamma_0$, $q$$\rangle$ where $\gamma_0$ $\in$ $\Gamma$, $q$ $\in$ \Q\ and checks if there exist a reachability plan $P$ such that $\gamma_0$ $\overset{P_\Psi}{\rightsquigarrow}$ $\gamma^\prime$ and $\gamma^\prime$ $\vdash_{\RPSAME}$ $q$.
\item \RPSUPER or relaxed reachability problem instance \I is of the form $\langle$$\gamma_0$, $q$$\rangle$ where $\gamma_0$ $\in$ $\Gamma$, $q$ $\in$ \Q\  and checks if there exist a reachability plan $P$ such that $\gamma_0$ $\overset{P_\Psi}{\rightsquigarrow}$ $\gamma^\prime$ and $\gamma^\prime$ $\vdash_{\RPSUPER}$ $q$.
\end{itemize}
\end{mydef}

\section{Pspace-complete reachability}\label{sec:pspace}
In this section, we present our reachability analysis results for different \rGURAG schemes shown in Figure \ref{fig:rgurag}. These results are extensions to the results from \GURA reachability analysis \cite{jin2013reachability} and also considers groups for assigning attributes to its member users. Our analysis will prove that \rGURAG schemes in Figure \ref{fig:rgurag} in general are \ps-complete. For such schemes we will first show that all \rGURAG schemes are in \ps~and then reduce a known \ps-complete problem to our problem schemes. In the next section, we will provide polynomial algorithms for some restricted \rGURAG problem classes.

\begin{lem} \label{lemma:pspace}
Reachability problem for every \rGURAG scheme in Figure \ref{fig:rgurag} is in PSPACE.
\end{lem}
\begin{proof}
Each state of a non-deterministic Turing machine stores some information to predict future states. This information takes polynomial amount of space and therefore all instance are in PSPACE. This proof is similarly stated for \GURA schemes in \cite{jin2013reachability} and more details are presented in appendix.
\end{proof}
Since all \rGURAG schemes are in \ps, it will now be sufficient to prove that all \rGURAG schemes are \ps-hard, which will conclude that the schemes are \ps-complete.

\begin{col}\label{col:gurag0}
Reachability query types \RPSUPER and \RPSAME for $\mathrm{\miniGURANG}$ schemes in general is
PSPACE-complete.
\end{col}
\begin{proof}
 Recall that Figure \ref{fig:rgurag} defines the relation between different \rGURAG schemes and \miniGURANZ. The reachability analysis for \miniGURANZ~scheme discussed in \cite{jin2013reachability} describes the scheme is PSPACE-complete. This scheme only allows change in attributes of the user. With respect to \miniGURANGZ, it can be said that \miniGURANZ~scheme is a sub-problem without user groups.  Therefore, the reduction from known PSPACE-complete problem (\miniGURANZ) to \miniGURANGZ is straightforward, which makes \miniGURANGZ as PSPACE-hard. Further, using Lemma \ref{lemma:pspace}, it is justified to claim that \miniGURANGZ is in PSPACE-complete.

Similar claim can also be made for \miniGURANGO scheme where \miniGURANGZ is its sub-problem involving only the same attribute in preconditions for rules ($\Psi$). Therefore, \miniGURANGO is PSPACE-hard and using Lemma \ref{lemma:pspace}, it is also PSPACE-complete. The analysis for \miniGURANGOP is also alike the above two schemes where \miniGURANGO is a sub-problem of \miniGURANGOP, therefore, \miniGURANGOP is in PSPACE-hard and hence PSPACE-complete also. More details are present in appendix.
\end{proof}

\section{Polynomial reachability for restricted cases} \label{s:proof}

In previous section, we proved that attribute reachability for any \rGURAG scheme in general is PSPACE-complete. However, we have identified some instances of \rGURAG schemes which can be solved in polynomial time under precondition restrictions on administrative rules ($\Psi$). Similar to \cite{jin2013reachability}, the following restrictions are considered where \D and \SR~are always imposed together:
\begin{itemize}
\item \textbf{No negation (\N)}: $\Psi$ satisfies \N
if no administrative rules in $\Psi$ use negation in preconditions.
\item \textbf{No deletion (\D)}: $\Psi$ satisfies \D
if for each attribute $att$ $\in$ \ATTR, $\mathsf{canDeleteU}_{att}$ and $\mathsf{canDeleteUG}_{att}$ are empty. Further, $\Remove$ rules are also empty, meaning, attribute values or groups once added cannot be deleted.
\item \textbf{Single rule with direct values (\SR)}: $\Psi$ satisfies \SR\xspace if for each attribute $att$ $\in$ \ATTR, there is at most one precondition associated with a particular value assignment in rules of $\mathsf{canAddU_{att}}$ or $\mathsf{canAddUG_{att}}$. Therefore, an attribute value pair can either be added through user directly or through groups but not both. Similar, condition also exists for $\Assign$ rules. Further, only direct conjuncts i.e. $val$ $\in$ $att_i(u)$, $val$ $\in$ $att_i(ug)$ or $ug$ $\in$ \directug($u$) are allowed in prerequisite condition.

\end{itemize}
\begin{algorithm*}[t]  \footnotesize
\centering
\caption{Plan Generation for \RPSAME in [\miniGURANGOP -- \N]}
\label{algo:sa-1+n}
\begin{algorithmic}[1]
\State{\textbf{Input:} problem instance I = $\langle$$\gamma_0$, $q$$\rangle$  \textbf{Output:} $plan$ or false}
\State{$plan$ := $\langle\rangle$;} \Comment{\texttt{Initialize plan}}
\State{$s$ := $\gamma_0$;} \Comment{\texttt{Initialize with state s}}
\State{\textbf{if} ($\exists$ $att$ $\in$ \ATTR $\exists$ $\mathrm{\langle}$$u$, $e\_{{att}}$, $vset$$\mathrm{\rangle}$ $\in$ $q$). $e\_att(u)$ $-$ $vset$ $\neq$ $\emptyset$ \textbf{then return} false;} 
\Comment{\texttt{Check if state s has more values than query}}
\LeftComment{\texttt{Assign attribute values required in query to the user or its effective groups}}
\While{($s$ $\nvdash_{\RPSAME}$ $q$ $\wedge$ \State{(($\exists$ $att'$ := $att$ $\in$ \ATTR $\exists$ $rule$ := $\mathsf{\langle}ar, c, val\mathrm{\rangle}$ $\in$ $\mathsf{canAddU}_{att'}$). ($\Evaluate_u(u,~c,~s)$ $\wedge$ $val$ $\notin$ $att'$($u$) $\wedge$ $\exists$ $\mathrm{\langle}$$u$, $e\_{{att'}}$, $vset$$\mathrm{\rangle}$ $\in$ $q$. $val$ $\in$ $vset$ ))
\State{$\lor$}
\State{(($\exists$ $att'$ := $att$ $\in$ \ATTR $\exists$ $rule$ := $\mathsf{\langle}ar, c, val\mathrm{\rangle}$ $\in$ $\mathsf{canAddUG}_{att'}$). ($\exists$ $ug'$ := $ug$ $\in$ $\effectiveug(u)$. $\Evaluate_{ug}(ug',~c,~s)$ $\wedge$ $val$ $\notin$ $att'$($ug'$) $\wedge$ } \State{$\hspace{112 mm}$ $\exists$ $\mathrm{\langle}$$u$, $e\_{{att'}}$, $vset$$\mathrm{\rangle}$ $\in$ $q$. $val$ $\in$ $vset$))}
  \State{$\lor$}
  \State{(($\exists$ $ug''$ := $ug$ $\in$ \UG $\exists$ $rule$ := $\mathsf{\langle}ar, c, ug''\mathrm{\rangle}$ $\in$ $\Assign$). ($\Evaluate_{u{-}ug}(u,~c,~s)$ $\wedge$ $ug''$ $\notin$ $\directug(u)$ $\wedge$} \State{$\hspace{85 mm}$ $\forall$ $att$ $\in$ \ATTR~$\exists$$\mathrm{\langle}$$u$, $e\_{{att}}$, $vset$$\mathrm{\rangle}$ $\in$ $q$. $e\_att(ug'')$ $\subseteq$ $vset$)))}}}
   \State{$\hspace{4 mm}$$s$ := $s$ $\ll$ $rule$; \Comment{\texttt{apply rule on state s}}}
   \State{\textbf{switch}}
   \Comment{\texttt{append administrative request to plan}}
   \State{$\hspace{4 mm}$\textbf{case} $rule$ $\in$ $\mathsf{canAddU}_{att'}$: \State{$\hspace{8 mm}$ $plan$ := $plan$.$\mathbf{append}$($\mathsf{add}(ar, u, att', val)$);}}
   \State{$\hspace{4 mm}$\textbf{break};}
   \State{$\hspace{4 mm}$\textbf{case} $rule$ $\in$ $\mathsf{canAddUG}_{att'}$: \State{$\hspace{8 mm}$ $plan$ := $plan$.$\mathbf{append}$($\mathsf{add}(ar, ug', att', val)$);}}
   \State{$\hspace{4 mm}$\textbf{break};}
   \State{$\hspace{4 mm}$\textbf{case} $rule$ $\in$ $\Assign$:
   \State{$\hspace{8 mm}$ $plan$ := $plan$.$\mathbf{append}$($\mathsf{assign}(ar, u, ug'')$);}}
   \State{$\hspace{4 mm}$\textbf{break};}
\EndWhile
\State{\textbf{if} $s$ $\vdash_{\RPSAME}$ $q$ \textbf{then} \textbf{return}
 $plan$ \textbf{else return false end if} }  \Comment{\texttt{check if reachability query is satisfied}}

\end{algorithmic}
\end{algorithm*} 
These restrictions are important in different kinds of attributes and scenarios. For instance, \textbf{No negation (\N)} restrictions have significance when attributes like $course$ or $degree$ are added to entities. It is likely that adding a new value for $course$ attribute do not require negation of another course as the precondition. Similarly, \textbf{No deletion (\D)} restriction can apply for attribute like $skills$ where a value once added to any entity will never be deleted. The \SR~restriction allows only unique preconditions in administrative relations for user and user groups. This restriction essentially separates set of attributes into two parts, one which can be assigned only to user directly and others assigned through groups. For example, attribute like $roomAccess$ can be assigned through group as it is usually common to all users with certain characteristics, and if value changes for one user, it will change for all others too. Attribute like $advisor$ is assigned individually to each user as change for one user may not change it in others. Therefore, these restrictions are relevant in applications.
\begin{algorithm*}[] \footnotesize
\centering
\caption{Group Assignment Plan Generation for \RPSAME in [\miniGURANGOP -- \D, \SR]}
\label{algo:sa-1+sr}
\begin{algorithmic}[1]
\State{\textbf{Input:} problem instance I = $\langle$$\gamma_0$, $q$$\rangle$ \textbf{Output:} $plan_{ug}$}
\State{\textbf{if} $\gamma_0$ $\vdash_{\RPSAME}$ $q$ \textbf{then} \textbf{return} $plan_{ug}$  := $\langle\rangle$;}
\Comment{\texttt{Check initial state}}
\State{$G_{ug} := \langle V_{ug}, E_{ug}\rangle$; $V_{ug}$ := \{$ug$ $|$ $\exists$ $ug$ $\in$ \UG. $\exists$ $ug$ $\notin$ $\directug(u)$. $\exists$ $\mathrm{\langle}ar, c, ug\mathrm{\rangle}$ $\in$ $\Assign(u)$. $\forall$ $att$ $\in$ \ATTR~$\exists$$\mathrm{\langle}$$u$, $e\_{{att}}$, $vset$$\mathrm{\rangle}$ $\in$ $q$. $e\_att(ug)$ $\subseteq$ $vset$ \}; $E_{ug}$ := $\emptyset$;}\Comment{\texttt{Construct a directed graph}}
\For{each pair of nodes ($(ug_1, ug_2) \in V_{ug})$} \State{\textbf{if} (($\exists \mathrm{\langle}ar, c, ug_2\mathrm{\rangle} \in$ $\mathsf{\Assign}$. ``$(ug_1 \in \directug(u))$" is a conjunct in $c$) $\lor$ \State{($\exists \mathrm{\langle}ar, c, ug_1\mathrm{\rangle} \in$
 $\mathsf{\Assign}$. ``$\neg (ug_2 \in \directug(u))$" is a conjunct in $c$))}}
 \State{\textbf{then} $E_{ug}$ := $E_{ug}$ $\cup$ $\{\langle ug_1, ug_2 \rangle\};$ \textbf{end if}} \Comment{\texttt{Add edges}}
\EndFor
\State{\textbf{if} graph $G_{ug}$ has cycles \textbf{then} remove the cyclic paths and $plan_{ug}$ :=  sequence of $\mathsf{assign} $ requests corresponding to the topological sort of $G_{ug}$;}

\end{algorithmic}
\end{algorithm*} 
\begin{algorithm*}[t] \footnotesize
\centering
\caption{Plan Generation for \RPSAME in [\miniGURANGO -- \D, \SR]}
\label{algo:sa-1sr}
\begin{algorithmic}[1]
\State{\textbf{Input:} problem instance I = $\langle$$\gamma_0$, $q$$\rangle$
\textbf{Output:} $plan$ or false}
\State{\textbf{if} $\gamma_0$ $\vdash_{\RPSAME}$ $q$ \textbf{then} \textbf{return} $plan$ := $\langle\rangle$;}
\Comment{\texttt{Check initial state}}
\State{$toadd$ := \{($att$, $val$) $|$ $att$ $\in$ \ATTR, $\mathrm{\langle}$$u$, $e\_{{att}}$, $vset$$\mathrm{\rangle}$ $\in$ $q$, $val$ $\in$ $vset$ \}}\Comment{\texttt{Values required in query}}
\State{$cur_{u}$ := \{($att$, $val$) $|$ $att$ $\in$ \ATTR, $val$ $\in$ $att(u)$\}}\Comment{\texttt{Current values of user}}
\State{\textbf{for} each $ug$ $\in$ $\effectiveug($u$)$ \textbf{do} $cur_{ug}$ := \{($att$, $val$) $|$ $att$ $\in$ \ATTR, $val$ $\in$ $att(ug)$\} \textbf{end for}}\Comment{\texttt{Current values of user's effective groups}}
\State{\textbf{if} ($cur_u$ $\cup$ $(\bigcup\limits_{ug \; \in\; \effectiveug(u)}cur_{ug}))$ $-$ $toadd$ $\neq$ $\emptyset$ \textbf{then return} false;}\Comment{\texttt{Check if state $\gamma_0$ has more values than query}}
\State{$ppre_{u}$ := $\emptyset$; \textbf{for} each $ug \in \effectiveug(u)$ \textbf{do} $ppre_{ug}$ := $\emptyset$; \textbf{end for}}
\For{(each $(att, val)$ $\in$ $toadd$ $\cup$ $ppre_u$)}
\Comment{\texttt{Positive precondition values for user}}
\State{$\hspace{2 mm}$ $ppre'_u$ := \{$(att_1, val_1)$ $|$ $\exists$ $\mathrm{\langle}ar, c, val
\mathrm{\rangle}$ $\in$ $\mathsf{canAddU}_{att}$. ``$val_1\in att_1(u)$" is a conjunct in $c$\};}
\State{$\hspace{2 mm}$ $ppre_u$ := ( $ppre_u$ $\cup$ ($ppre'_u$ $\backslash$ $ppre_u$ )) $\backslash$ $cur_{u}$;}
\EndFor
\For{(each $ug \in \effectiveug(u)$)}
\Comment{\texttt{Positive precondition values for effective groups}}
\For{(each $(att, val)$ $\in$ $toadd$ $\cup$ $ppre_{ug}$)}
\State{$ppre'_{ug}$ := \{$(att_1, val_1)$ $|$ $\exists$ $\mathrm{\langle}ar, c, val
\mathrm{\rangle}$ $\in$ $\mathsf{canAddUG}_{att}$. ``$val_1\in att_1(ug)$" is a conjunct in $c$\};}
\State{$ppre_{ug}$ := ( $ppre_{ug}$ $\cup$  ($ppre'_{ug}$ $\backslash$ $ppre_{ug}$)) $\backslash$ $cur_{ug}$;}
\EndFor
\EndFor
\LeftComment{\texttt{ Check if rules exists for values required }}
\State{\textbf{if} (($\exists(att, val)$ $\in$ $toadd$ $\cup$ $ppre_u$ $\cup$ $(\bigcup\limits_{ug \; \in\; \effectiveug(u)}ppre_{ug})$  $\backslash$ ($cur_u$ $\cup$ $(\bigcup\limits_{ug \; \in\; \effectiveug(u)} cur_{ug})))$. $\nexists$ $\mathrm{\langle}ar, c, val \mathrm{\rangle}$ $\in$ $\mathsf{canAddU}_{att}$ $\cup$ $\mathsf{canAddUG}_{att}$)\textbf{then return} false;}
\LeftComment{\texttt{Find negation values in rules required to add values for the user and its effective groups}}
\State{$npre_u$ := \{($att_1$, $val_1$) $|$ $\exists$ $(att, val)$ $\in$ ($toadd$ $\cup$ $ppre_u$) $\backslash$ $cur_u$. $\exists$ $\mathrm{\langle}ar, c, val
\mathrm{\rangle}$ $\in$ $\mathsf{canAddU}_{att}$. ``$\neg (val_1\in att_1(u))$" is a conjunct in $c$\}}
\State{\textbf{for} each $ug \in \effectiveug(u)$ \textbf{do} $npre_{ug}$ := \{($att_1$, $val_1$) $|$ $\exists$ $(att, val)$ $\in$ ($toadd$ $\cup$ $ppre_{ug}$) $\backslash$ $cur_{ug}$ . $\exists$ $\mathrm{\langle}ar, c, val
\mathrm{\rangle}$ $\in$ $\mathsf{canAddUG}_{att}$. ``$\neg (val_1\in att_1(ug))$" is a conjunct in $c$\} \textbf{end for}}
\State{\textbf{if} (($npre_u$ $\cap$ $cur_{u}$ $\neq$ $\emptyset$) $\lor$ ($\forall$ $ug$ $\in$ \effectiveug($u$). $npre_{ug}$ $\cap$ $cur_{ug}$ $\neq$ $\emptyset$)) \textbf{then return} false;}
\Comment{\texttt{Negation in current values}}
\State{$G := \langle V, E\rangle$; $V$ := $toadd$ $\cup$ $ppre_{u}$ $\cup$ $(\bigcup\limits_{ug \; \in\; \effectiveug(u)}ppre_{ug})$ $\backslash$ ($cur_u$ $\cup$ $(\bigcup\limits_{ug \; \in\; \effectiveug(u)} cur_{ug}))$; $E$ := $\emptyset$;}\Comment{\texttt{Construct a directed graph}}
\For{each pair of nodes ($(att_1, val_1),(att_2, val_2)$) $\in$ $V$}
\State{\textbf{if} ((($\exists \mathrm{\langle}ar, c, val_2\mathrm{\rangle} \in$ $\mathsf{canAddU}_{att_2}$. ``$(val_1 \in att_1(u))$" is a conjunct in $c$) $\lor$ \State{($\exists \mathrm{\langle}ar, c, val_1\mathrm{\rangle} \in$
 $\mathsf{canAddU}_{att_1}$. ``$\neg (val_2 \in att_2(u))$" is a conjunct in $c$))}}
 \State{ $\lor$}
 \State{(($\exists$ $ug \in \effectiveug(u)$). (($\exists \mathrm{\langle}ar, c, val_2\mathrm{\rangle} \in \mathsf{canAddUG}_{att_2}$. ``$(val_1 \in att_1(ug))$" is a conjunct in $c$) $\lor$ \State{($\exists \mathrm{\langle}ar, c, val_1\mathrm{\rangle} \in$ $\mathsf{canAddUG}_{att_1}$. ``$\neg (val_2 \in att_2(ug))$" is a conjunct in $c$))))}}
\State{\textbf{then} $E := E$ $\cup$ \{$\langle(att_1, val_1),(att_2, val_2)\rangle$\}; \textbf{end if} }\Comment{\texttt{Add edges to the graph}}
\EndFor
\State{$valset$ := $toadd$ $-$ ($cur_{u}$ $\cup$ $(\bigcup\limits_{ug \; \in\; \effectiveug(u)}cur_{ug})$ );}
\Comment{\texttt{Values in query not in state $\gamma$}}
\State{\textbf{if} $\exists (att_1,val_1) \in valset~\exists \langle(att, val),(att_1, val_1)\rangle \in E$. $(att, val) \notin valset$ \textbf{then return} false \State{\textbf{else} $V := vset$ $E := E - \{\langle(att, val),(att_1, val_1)\rangle$ $|$ $(att, val)$ $\notin$ $vset$, $(att_1, val_1)$ $\notin$ $valset$\} \textbf{end if}}}
\State{\textbf{if} graph G has a cycle \textbf{then return} false \textbf{else return} $plan$ := sequence of administrative requests corresponding to the topological sort of G;}
\end{algorithmic}
\end{algorithm*} 

We now discuss reachability analysis for restricted \rGURAG schemes. The notation [\rGURAGN$_x$, \texttt{Restriction}] specifies special instances of \rGURAG scheme where subscript $x$ takes a value in 0, 1 or 1+ representing -- \miniGURANGZ, \miniGURANGO or \miniGURANGT and {\tt Restriction} represents combinations of \N, \D and \SR~specifying that administrative rules $\Psi$ in the scheme satisfy these restrictions. For example, [\miniGURANGZ -- \N] denotes \miniGURANGZ scheme where rules in $\Psi$ satisfy \N.

As shown in Figure \ref{fig:rgurag}, \miniGURANGOP scheme is the most expressive scheme where new attribute values are achieved by direct assignment to the user or to its effective groups, and also by changing user to group memberships. It is clear from the previous discussions that the scheme covers \miniGURANGO and \miniGURANGZ, which only allow change in attributes of the user or its effective groups. Therefore, we will only discuss algorithm for restricted \miniGURANGOP scheme which can be easily used for other two schemes by simply ignoring irrelevant administrative rules.
\subsection{Reachability plan for RP\textsubscript{=} in [\upshape{r}GURA\textsubscript{G\textsubscript{1+}}{--} $\mathbf{\overline{N}}$] }
First we will discuss reachability query type \RPSAME for $\mathrm{[\miniGURANGOP{-}\overline{N}]}$ scheme which can be solved in polynomial time by Algorithm \ref{algo:sa-1+n}. This algorithm extends the algorithm discussed for \miniGURANO~\cite{jin2013reachability} by including user group attribute assignments and also modification in user to group memberships. The added restriction to this scheme (\N) requires preconditions in rules without negation conjuncts and therefore, administrative rules cannot specify addition of new attributes based on the absence of some other values. Hence, the current attribute values of user or groups are not required to be removed for adding new values or group, which precludes the need for investigating any $\canD_{att}$, $\canDUG_{att}$ and $\Remove$ rules.

The algorithm starts with the current set of attribute values and group memberships for user, and the attribute values for its member groups. It traverse all relevant $\canA_{att}$, $\canAUG_{att}$ or $\Assign$ rules to add new values to the attributes of user or to its effective user groups and also add new groups to the user. Since, the query type is restricted, the algorithm first checks if the current effective attributes of user are not more than what required in the query (line 4). If the current values are extra, the algorithm returns false, since there are no delete administrative relations to delete such values. The while loop (line 5--24) terminates when either the query is satisfied or when no other values can be added from the rules in $\canA_{att}$ and $\canAUG_{att}$ or no new groups can be added to the user using $\Assign$ rules. When adding a new value to the user or its effective groups, the corresponding value must be checked against the query. If the value is present in the query, the addition is allowed. Similar check is also done to add new groups the user, where all the attributes present in the group should also be a part of the query. The order to add these values or new groups is independent to each other, since no negation conjuncts are required and presence of extra values in user or group will not stop from adding new values. Also, if later a new value is added to an entity, the while loop will again consider the relevant rules to add values based on the already added values. When a new attribute is added to user, its effective groups or a new group is assigned to the user, its corresponding administrative request is appended to the reachability plan $plan$. If the query is satisfied, the algorithm returns the corresponding reachability plan $plan$ or returns false stating that the query is unsatisfiable and user will not achieve desired effective attributes as mentioned in query.

\begin{thm}\label{thm-sa-1+n}
Reachability query type \RPSAME for scheme $\mathrm{[\miniGURANGOP- \overline{N}]}$ is P.
\end{thm}
\begin{proof}
Algorithm \ref{algo:sa-1+n} describes the polynomial time algorithm.

\noindent
\textbf{Complexity:} The complexity is determined by the number of times the administrative rules in $\canA_{att}$, $\canAUG_{att}$ or $\Assign$ are traversed. If only one value is added by each of the rules, the complexity of Algorithm \ref{algo:sa-1+n} is $\mathcal{O}$($|\mathsf{\Assign}|$ $\times$ $|\UG|$ + (($\sum_{att \in \ATTR}|\SCOPE_{att}|$) $\times$ ( $|\mathsf{\canA_{att}}|$ + $|\mathsf{\canAUG_{att}}|$ $\times$ $|\UG|$ ))), where $|\canA_{att}|$, $|\canAUG_{att}|$ and $|\Assign|$ represents number of the administrative rules in these relations and $|\UG|$ represents the maximum number of groups assigned to the user. Clearly, the complexity of algorithm is polynomial.
\end{proof}

The \RPSUPER query type for $\mathrm{[\miniGURANGOP- \overline{N}]}$ also has a polynomial algorithm, where the extra conditions to check the query before adding new values is removed since we can have values even if they are not required in the query. The complexity will remain the same as shown in Theorem \ref{thm-sa-1+n}.
Similar algorithm can also be devised for \RPSAME and \RPSUPER query type in $\mathrm{[\miniGURANGO{-}\overline{N}]}$ and $\mathrm{[\miniGURANGZ{-}\overline{N}]}$ schemes where $\Assign$ rules will not be considered into the while loop for adding new groups to the user. Hence these schemes can be also solved in polynomial time.

\subsection{Reachability plan for RP\textsubscript{=} in [\upshape{r}GURA\textsubscript{G\textsubscript{1+}}{--} $\mathbf{\overline{D}, SR\textsubscript{d}}$]}
We will now consider another restricted instance for \miniGURANGOP, $\mathrm{[\miniGURANGOP- \overline{D},\SR]}$ which can be solved by Algorithm \ref{algo:sa-1+sr} and \ref{algo:sa-1sr}. The scheme has two restrictions, $\mathrm{\overline{D}}$ which removes the need to consider delete administrative relations -- $\canD_{att}$,  $\canDUG_{att}$ and $\Remove$. The \SR~restriction allows single preconditions for each attribute value pair or user group, with only direct values as conjuncts in preconditions. This restriction results in rules which can be either satisfied by user or any of its effective groups but not both. We have divided the algorithm into two algorithm for ease of understanding and to show how these algorithms can be reused in other schemes also.

Algorithm \ref{algo:sa-1+sr} is used to add new groups to the user. Since the preconditions only involves user's direct groups as conjuncts (\SR~restriction), the addition of groups is independent of the attributes and can be calculated separately. The administrative rules in this scheme can have negation conjuncts in preconditions, therefore, the order of assigning new groups can be mutually dependent. The algorithm first creates a directed graph where vertices $V_{ug}$ are user groups and edges $E_{ug}$ are directed based on conjuncts in precondition of rules in $\Assign$. In line 3, before adding a group to $V_{ug}$, it is checked that all the attributes in group are required in query, as no extra attributes are allowed in \RPSAME and deletion is not allowed. Line 4 -- 8 creates edges in the graph, if a user group $ug_1$ is a negation conjunct to add another group $ug_2$ or $ug_2$ is a precondition for $ug_1$, then edge is drawn from $ug_2$ to $ug_1$, signifying that $ug_2$ should be added before $ug_1$. If cycles exists in the created graph then remove the cyclic paths and create topological sort on the remaining graph. The set of administrative requests based on the sort will provide the $plan_{ug}$ for user to groups assignment. Once the requests are executed in order, new effective groups are calculated for the user and computation continues from Algorithm \ref{algo:sa-1sr}.

Algorithm \ref{algo:sa-1sr} extends algorithm defined in \cite{jin2013reachability}, which checks the final set of values required to satisfy the reachability query and find $\canA_{att}$ or $\canAUG_{att}$ rules to add those values. Further to add the values in precondition of rules, it may in-turn need some other rules and values and so on. Therefore, algorithm traverses backward to find the set of values required to satisfy the query. Since the values can be achieved by user directly or from any of its effective groups, this backward search is done for user and all its effective groups as calculated by Algorithm \ref{algo:sa-1+sr}.
\begin{table*}[t]
\centering
\caption{Example Problem Instance for \RPSAME in [\miniGURANGOP -- \N]}
\label{tab:ex-1}
{%
\begin{tabular}{llllll}
\toprule
\multicolumn{6}{l}{\;\; \textbf{Input:} problem instance I = $\langle$$\gamma_0$, $q$$\rangle$
\textbf{Output:} $plan$ or false}\\
\multicolumn{6}{l}{\begin{tabular}[c]{@{}l@{}}\;\; $\psi \in \Psi$ : \\\;\; $\mathrm{ \mathsf{canAddU}_{roomAcc}}$ = \{$\mathrm{\langle BuildAdmin,~ c{++} \in e\_skills(\mathit{u}) \wedge 2.04 \in roomAcc(\mathit{u}),~ 1.2 \rangle}$ \},
\\\;\; $\mathrm{\mathsf{canAddU}_{college} = \{\langle BuildAdmin,~ python \in e\_skills(\mathit{u}) \wedge 3.05 \in roomAcc(\mathit{u}),~ COE \rangle}\}$,
\\\;\; $\mathrm{\mathsf{canAddU}_{skills} = \{\langle DeptAdmin,~ c \in e\_skills(\mathit{u}),~ python \rangle}\}$,
\\\;\; $\mathrm{ {\mathsf{canAddUG}_{roomAcc}} = \{\langle BuildAdmin,~ 3.02 \in roomAcc(\mathit{ug}),~ 1.2 \rangle }$ \},
\\\;\; $\mathrm{ \mathsf{\Assign} = \{\langle DeptAdmin,~ G_1 \in \directug(\mathit{u}),~ G_3 \rangle }$ \}  \end{tabular}}\\
\multicolumn{6}{l}{\;\; \textbf{Queries:}}\\
\multicolumn{6}{l}{\begin{tabular}[c]{@{}l@{}}\;\; $q_1$ $\in$ \Q~= \;\; \{$\mathrm{e\_roomAcc(u) = \{2.04, 2.03, 3.02, 1.2\}, \;\; e\_skills(u) = \{c, c{++}, python\}, \;\; e\_college(u) = \{COS\}}$\} \end{tabular}}\\
\multicolumn{6}{l}{\begin{tabular}[c]{@{}l@{}}\;\; $q_2$ $\in$ \Q~= \;\; \{$\mathrm{e\_roomAcc(u) = \{2.04, 2.03, 3.02, 1.2\}, \;\; e\_skills(u) = \{c, c{++}\}, \;\; e\_college(u) = \{COS, COE\}}$\} \end{tabular}}\\
\bottomrule
\end{tabular}
}
\end{table*}

The algorithm starts by checking if the query is satisfied in the current state, in that case empty plan is returned signifying that with only new group assignments query is satisfied. Otherwise, it creates a set of attribute value pair for values required in query $q$ and also for current attributes of user and its effective groups (line 4-5). Line 6 checks if the union of current values of the user or its effective groups is not more than values required in $q$. If extra values are there, the algorithm returns false, as no delete rules are allowed. The algorithm calculates all positive precondition attribute value pairs required by user or its effective groups to get values in $toadd$ (line 7-17). Therefore, the final set of values required includes the values in query ($toadd$) and positive preconditions in user or its effective groups excluding their current values. Line 18 checks if rules exists to add all required attribute value pair or else returns false, as the values can not be added. Line 19-21 calculate negative conjuncts in rules required to add required values and returns false if the such values are present in current state. After passing through all checks, the algorithm starts creating a directed graph. Vertices ($V$) in the graph are attribute value pair of the values required in the query $q$ and the required positive preconditions excluding the values in the current state. Edges $E$ will be drawn in the direction defined in the for loop (line 23-30). If the attribute value pair $(att_1,val_1)$ is in the negative conjunct in administrative rule for $(att_2,val_2)$ or $(att_2,val_2)$ is a positive conjunct in a rule to add $(att_1,val_1)$, the edge is created from $(att_2,val_2)$ to $(att_1,val_1)$. Since our query type is \RPSAME, it requires an additional check so that no extra values are added to the user. Therefore, once the graph is created, we create a set $valset$, which includes values required in query and not present in the current state. If the created graph has vertex in $valset$ having incoming edge not from vertex in $valset$, algorithm returns false (line 32). Otherwise it removes all the edges from vertices not in $valset$. If cycles exists in the remaining graph then algorithm returns false, else the set of administrative request corresponding to the topological sort will return the $plan$.

Therefore, the overall reachability plan returned will be $plan_{ug}$ from Algorithm \ref{algo:sa-1+sr} and $plan$ from Algorithm \ref{algo:sa-1sr}.

\begin{thm}\label{thm-sa-1+sr}
Reachability query type \RPSAME for scheme $\mathrm{[\miniGURANGOP{-}\overline{D},\SR]}$ is P.
\end{thm}
\begin{proof}
Algorithm \ref{algo:sa-1+sr} and \ref{algo:sa-1sr} describe the polynomial algorithms.

\noindent
\textbf{Complexity:} The algorithm takes polynomial time to create directed graphs and then to compute topological sort. Its complexity is $\mathcal{O}$($|\UG|$ $\times$ $|\mathsf{\Assign}|$ + (($\sum_{att \in \ATTR}|\SCOPE_{att}|$) $\times$ ($|\mathsf{\canA_{att}}|$ + $|\canAUG_{att}|$ $\times$ $|\UG|$))).
\end{proof}
In case of \RPSUPER query type for $\mathrm{[\miniGURANGOP{-}\overline{D},\SR]}$, we remove the extra checks to verify if no extra values are present in current state (line 6). Further line 31-33 is not required as extra values are allowed to be added to user. With these minor changes, the complexity of \RPSUPER for scheme $\mathrm{[\miniGURANGOP{-}\overline{D},\SR]}$ is P.

It should be noted that \RPSAME for $\mathrm{[\miniGURANGO{-}\overline{D},\SR]}$ do no allow changes in group memberships of user. Therefore, computation for this scheme will start directly from Algorithm \ref{algo:sa-1sr}, obviating the execution of Algorithm \ref{algo:sa-1+sr}. The \RPSUPER query for $\mathrm{[\miniGURANGO{-}\overline{D},\SR]}$ will remove all extra conditions applied in Algorithm \ref{algo:sa-1sr} for \RPSUPER scheme for $\mathrm{[\miniGURANGOP{-}\overline{D},\SR]}$ as discussed above. Also, since \miniGURANGZ is a sub-problem of \miniGURANGO we can conjecture that \RPSAME and \RPSUPER for scheme $\mathrm{[\miniGURANGZ{-}\overline{D},\SR]}$ can be solved in polynomial time.
 
\section{Example Problem Instance }
\label{sec:example}

\begin{figure}[t]
\centering
\includegraphics[scale=.55]{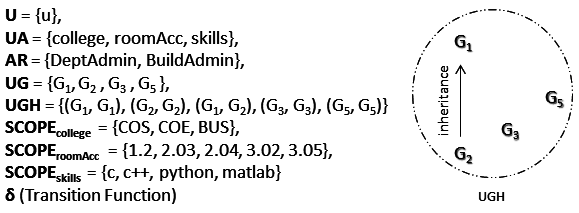}
\caption{Input Starting State ($\mathrm{\gamma_0 \in \Gamma}$)}
\label{fig:input}
\end{figure}

We will now illustrate the plan generation in two schemes discussed earlier with a sample input state and a set of reachability queries. Figure \ref{fig:input} defines the common input for both the schemes.

\begin{figure}[t]
\centering
\includegraphics[scale=.60]{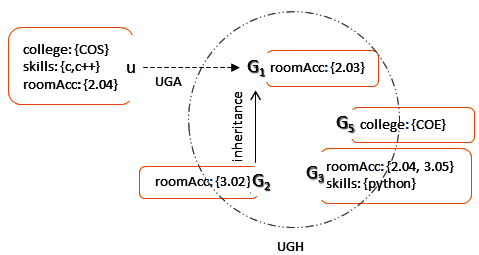}
\caption{Initial State for \RPSAME in [\miniGURANGOP -- \N]}
\label{fig:ex-1}
\end{figure}

\textbf{Plan Generation for \RPSAME in [\miniGURANGOP -- \N]}:
Figure \ref{fig:ex-1} shows attributes of user and groups along with user to group direct membership. Table \ref{tab:ex-1} defines set of administrative rules allowed in scheme along with two reachability queries. We will first try to find a reachability plan (if exists) for query $q_1$ using Algorithm \ref{algo:sa-1+n}.

Initially, $plan$ is set to empty $\langle$$\rangle$. The initial state is checked to find if it has more attribute values than required in query $q_1$. In state $\gamma_0$, the effective values of user are e\_roomAcc(u) = \{2.04, 2.03, 3.02\}, e\_skills(u) = \{c, c++\}, e\_college(u) = \{COS\}, which are all required in query. The while loop checks if query $q_1$ is satisfied in state $\gamma_0$, which is not true as some values are missing. Now algorithm starts adding new values to the user or its effective groups and also assign new groups to user based on administrative rules defined in Table \ref{tab:ex-1}. The first rule requires effective skills of user having value c++ and roomAcc attribute with value 2.04. to add 1.2 value to roomAcc by administrative role BuildAdmin. Since user satisfy these conditions and 1.2 value is not directly assigned in roomAcc(u) and the value is required in $q_1$, it adds 1.2 value to roomAcc(u). The administrative request $\mathsf{add}$( BuildAdmin, u, roomAcc, 1.2) is also appended to the $plan$. The algorithm again goes through the while loop and checks if $q_1$ is satisfied. The user is still missing skills attribute value python. The algorithm now tries to add group G$_3$ to user u. The precondition of $\Assign$ rule is satisfied by user, but the effective values for roomAcc attribute for group G$_3$ are \{3.05, 2.04\}, which is not the subset of values required in query. Hence, G$_3$ cannot be assigned to user u. Next, the algorithm checks rule for skills attribute to add value python and finds that preconditions to add value python are satisfied by user u. It appends the corresponding request $\mathsf{add}$( DeptAdmin, u, skills, python) to $plan$ which results in total of two requests in the plan. The algorithm again checks the new state against $q_1$ and finds the query is ``strict" satisfied. It breaks the while loop and returns plan = $\mathsf{add}$( BuildAdmin, u, roomAcc, 1.2), $\mathsf{add}$( DeptAdmin, u, skills, python).

We now check the satisfiability of query $q_2$ with the same initial state. Similar to $q_1$, query is checked against initial state to check extra values and value 1.2 for attribute roomAcc is added to the user and requests is appended to the plan. Second rule allows to add COE for attribute college, but the preconditions are not satisfied by user. We try to add group G$_3$ but it also adds extra values which are not required in query. It can be noticed that after all the administrative rules are checked, the query cannot be satisfied and hence the algorithm returns false.

\begin{figure}[t]
\centering
\includegraphics[scale=.60]{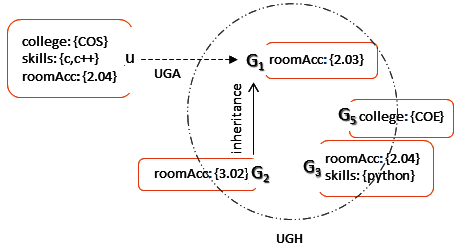}
\caption{Initial State for \RPSAME in [\miniGURANGOP -- \D, \SR]}
\label{fig:ex-2}
\end{figure}

\textbf{Plan Generation for \RPSAME in [\miniGURANGOP -- \D, \SR]}: Figure \ref{fig:ex-2} shows user and group attributes along with user to group direct membership. Table \ref{tab:ex-2} defines the set of administrative rules allowed in the scheme and three reachability queries. It should be noted that the rules in $\Psi$ have negation conjuncts and single precondition with direct attributes or group memberships for each attribute value pair or user group.
We will start with Algorithm \ref{algo:sa-1+sr} to assign new groups to the user. Once groups are assigned, attributes will be added to user or its newly computed effective groups. If Algorithm \ref{algo:sa-1+sr} doesn't add new groups, the computation will still be done by Algorithm \ref{algo:sa-1sr} with old effective groups.
\begin{table*}[t]
\centering
\caption{Example Problem Instance for \RPSAME in [\miniGURANGOP -- \D, \SR]}
\label{tab:ex-2}
{%
\begin{tabular}{llllll}
\toprule
\multicolumn{6}{l}{\textbf{Input:} problem instance I = $\langle$$\gamma_0$, $q$$\rangle$
\textbf{Output:} $plan$ or false}\\ 
\multicolumn{6}{l}{\begin{tabular}[c]{@{}l@{}}$\psi \in \Psi$ : \\ $\mathrm{\mathsf{canAddU}_{roomAcc} = \{\langle BuildAdmin,~ c{++} \in skills(\mathit{u}) \wedge \neg(2.04 \in roomAcc(\mathit{u})),~ 1.2 \rangle}$\}, \\$\mathrm{ \mathsf{canAddU}_{skills} = \{\langle DeptAdmin,~ c \in skills(\mathit{u}),~ python \rangle }$\},  \\
$\mathrm{\mathsf{canAddU}_{college} = \{\langle BuildAdmin,~ matlab \in skills(\mathit{u}),~ BUS \rangle}$\}, \\
$\mathrm{\mathsf{canAddU}_{skills} = \{\langle DeptAdmin,~ c \in skills(\mathit{u}) \wedge COS \in college(\mathit{u}),~ matlab \rangle}$\} \\
$\mathrm{\mathsf{canAddUG}_{college} = \{\langle BuildAdmin,~ python \in skills(\mathit{ug}) \wedge \neg(2.04 \in roomAcc(\mathit{ug})),~ COE \rangle}$\}, \\
$\mathrm{\mathsf{\Assign} = \{\langle DeptAdmin,~ G_1 \in \directug(\mathit{u}),~ G_3 \rangle}$,
$\mathrm{\langle DeptAdmin,~ \neg(G_3 \in \directug(\mathit{u})),~ G_5 \rangle}$\}   \end{tabular}}\\
\multicolumn{6}{l}{\textbf{Queries:}}\\
\multicolumn{6}{l}{\begin{tabular}[c]{@{}l@{}}$q_1$ $\in$ \Q~= \{$\mathrm{e\_roomAcc(u) = \{2.04, 2.03, 3.02\}, \;\; e\_skills(u) = \{c, c{++}, python\}, \;\; e\_college(u) = \{COS, COE\}}$\} \end{tabular}}\\
\multicolumn{6}{l}{\begin{tabular}[c]{@{}l@{}}$q_2$ $\in$ \Q~= \{$\mathrm{e\_roomAcc(u) = \{2.04, 2.03, 3.02, 1.2\}, \;\; e\_skills(u) = \{c, c{++}, python\}, \;\; e\_college(u) = \{COS, COE\}}$\} \end{tabular}}\\

\multicolumn{6}{l}{\begin{tabular}[c]{@{}l@{}}$q_3$ $\in$ \Q~= \{$\mathrm{e\_roomAcc(u) = \{2.04, 2.03, 3.02\}, \;\; e\_skills(u) = \{c, c{++}, python, matlab\}, \;\; e\_college(u) = \{COS, COE, BUS\}}$\} \end{tabular}}\\
\bottomrule
\end{tabular}
}
\end{table*}

Algorithm \ref{algo:sa-1+sr} creates group assignment plan (defined as $plan_{ug}$) to assign new groups to user. Two administrative rules exists in $\Assign$ relation. Since G$_3$ is not directly assigned to user u, precondition is satisfied and G$_3$ has value python for skill attribute, which is required in query $q_1$, algorithm adds G$_3$ to the set of vertices $V_{ug}$. Similarly group G$_5$ is also added to $V_{ug}$. There are no more $\Assign$ rules, hence the algorithm starts adding edges to the graph. For (G$_3$, G$_5$) $\in$ $V_{ug}$, since G$_3$ is a negation conjunct in precondition to add G$_5$, therefore, directed edge is drawn from  G$_5$ to G$_3$. As here are no other relevant $\Assign$ rules and vertices pair, it breaks the loop and creates a topological sort of the graph. Sort will have \{G$_5$, G$_3$\} order and the corresponding plan $plan_{ug}$ := $\mathrm{{\mathsf{assign}(DeptAdmin,~u,~G_5)}}$, $\mathrm{{\mathsf{assign}(DeptAdmin,~u,~G_3)}}$ is returned. Before proceeding to Algorithm \ref{algo:sa-1sr}, the request in $plan_{ug}$ must be executed to get new effective groups of the user. Algorithm \ref{algo:sa-1sr} is used to assign attributes to user and newly computed effective groups (which will now have group G$_5$ and G$_3$ along with G$_1$ and G$_2$). It first checks if the query ($q_1$) is satisfied in the current state (line 2) which has new direct groups assigned using algorithm \ref{algo:sa-1+sr}. Clearly query $q_1$ is satisfied with new group assignments only, hence the reachability plan for group assignments $plan_{ug}$ is returned.

For queries $q_2$ and $q_3$, group assignment plan $plan_{ug}$ is created similarly as above. Therefore, we will follow algorithm \ref{algo:sa-1sr} with user's effective groups as G$_1$, G$_2$, G$_3$ and G$_5$. For $q_2$, current state do not have value 1.2 for roomAcc attribute. The algorithm first computes $toadd$, which is the set of attribute value pair in $q_2$:
\begin{itemize}[leftmargin= 0cm]
\item [] $toadd$ = $\mathrm{\{\langle {roomAcc},~ 2.04 \rangle\ , \langle {roomAcc},~  2.03 \rangle ,}$ \\ $\mathrm{\;\;\;\;\;\;\;\;\;\;\;\;\; \langle {roomAcc},~ 3.02 \rangle\ , \langle {roomAcc},~ 1.2 \rangle ,}$ \\ $\mathrm{\;\;\;\;\;\;\;\;\;\;\;\;\; \langle {skills},~ c \rangle , \langle {skills},~ c{++} \rangle , \langle {skills},~ python \rangle ,}$ \\$\mathrm{\;\;\;\;\;\;\;\;\;\;\;\;\;\langle {college},~ COS \rangle , \langle {college},~ COE \rangle\}}$
\end{itemize}

It then calculates the current attribute value pair for user and its effective groups (Line 4-5):
\begin{itemize}[leftmargin= 0cm]
\item [] $cur_\mathrm{u}$ = $\mathrm{\{\langle {roomAcc},~ 2.04 \rangle\ , \langle {skills},~ c \rangle , \langle {skills},~ c{++} \rangle ,}$ \\ $\mathrm{\;\;\;\;\;\;\;\;\;\;\;\;\;\langle {college},~ COS \rangle \}}$
\item [] $cur_\mathrm{G_1}$ = $\mathrm{\{\langle {roomAcc},~ 2.03 \rangle\}}$ \;\; $cur_\mathrm{G_2}$ = $\mathrm{\{\langle {roomAcc},~ 3.02 \rangle\}}$
\item [] $cur_\mathrm{G_3}$ = $\mathrm{\{\langle {roomAcc},~ 2.04 \rangle\ , \langle {skills},~ python \rangle \}}$ \;\; \\ $cur_\mathrm{G_5}$ = $\mathrm{\{\langle {college},~ COE \rangle \}}$
\end{itemize}
\noindent
The algorithm checks if the current attributes of user and its effective groups are not extra than the values required in the query. Clearly, for query $q_2$, no extra values are present in current state. The algorithm next computes the positive conjuncts in the preconditions required to add the values in $toadd$. It first calculates for each attribute value pair in $toadd$ and then recalculates for each positive preconditions attribute value pair also. For example, positive conjunct for user to add $\mathrm{\langle {roomAcc},~ 1.2 \rangle}$ in $toadd$ is $\mathrm{\langle {skills},~ c{++} \rangle}$ and for $\mathrm{\langle {skills},~ python \rangle}$ in $toadd$ is $\mathrm{\langle {skills},~ c \rangle}$. Therefore (using line 9), $ppre'_u$ := \{$\mathrm{\langle {skills},~ c{++} \rangle}$, $\mathrm{\langle {skills},~ c \rangle}$\}. It then recomputes $ppre_u$ by combining its values with newly computed $ppre'_u$ after removing values already present in $ppre_u$ or $cur_u$. In this case, no new value is added in $ppre_u$, as both the values in $ppre'_u$ are already present in $cur_u$. Similarly, the positive preconditions are calculated for each effective groups.

\begin{itemize}[leftmargin= 0cm]
\item [] $ppre_\mathrm{u}$ = $\mathrm{\{\}}$,\;\; $ppre_\mathrm{G_1}$ =  $ppre_\mathrm{G_2}$ = $ppre_\mathrm{G_5}$ = $\mathrm{\{\langle {skills},~ python \rangle\}}$
\item [] $ppre_\mathrm{G_3}$ = $\mathrm{\{\}}$
\end{itemize}

\noindent
Next, in line 18, the algorithm checks if rules exists for values required in $toadd$ and positive preconditions excluding the current values. Clearly, rule exists for $\mathrm{\langle {roomAcc},~ 1.2 \rangle}$ pair and all other values are already present in user or its effective groups. It then calculates negative conjuncts for user and its effective groups in preconditions to add values in $toadd$ and positive preconditions excluding current state. For user, $\mathrm{\langle {roomAcc},~ 1.2 \rangle}$ has negation conjunct $\mathrm{\langle {roomAcc},~ 2.04 \rangle}$ in $\mathrm{\mathsf{canAddU}_{roomAcc}}$. Remaining negation conjuncts are as follows:
\begin{itemize}[leftmargin= 0cm]
\item [] $npre_\mathrm{u}$ = $\mathrm{\{\langle {roomAcc},~ 2.04 \rangle\}}$
\item [] $npre_\mathrm{G_1}$ =  $npre_\mathrm{G_2}$ = $npre_\mathrm{G_3}$ = $\mathrm{\{\langle {roomAcc},~ 2.04 \rangle\}}$
\item [] $npre_\mathrm{G_5}$ = $\mathrm{\{\}}$
\end{itemize}

\noindent
Line 21 checks if the negation conjuncts exists in current values of either user or its effective groups. User has $\mathrm{\{\langle {roomAcc},~ 2.04 \rangle\}}$ pair in $cur_u$, therefore, roomAcc attribute cannot get value 1.2 required in $q_2$ since only single rule exists for user or groups. Hence, the algorithm returns false for query $q_2$. For query $q_3$, $toadd$ values are:
\begin{itemize}[leftmargin= 0cm]
\item [] $toadd$ = $\mathrm{\{\langle {roomAcc},~ 2.04 \rangle\ , \langle {roomAcc},~  2.03 \rangle ,}$ \\ $\mathrm{\;\;\;\;\;\;\;\;\;\;\;\;\; \langle {roomAcc},~ 3.02 \rangle\ , \langle {skills},~ c \rangle , \langle {skills},~ c{++} \rangle ,}$ \\ $\mathrm{\;\;\;\;\;\;\;\;\;\;\;\;\; \langle {skills},~ python \rangle , \langle {skills},~ matlab \rangle}$ \\$\mathrm{\;\;\;\;\;\;\;\;\;\;\;\;\;\langle {college},~ COS \rangle , \langle {college},~ COE \rangle , \langle {college},~ BUS \rangle\}}$
\end{itemize}

\noindent
The current values are still the same as defined in query $q_2$. The algorithm calculates the positive conjuncts in preconditions as:
\begin{itemize}[leftmargin= 0cm]
\item [] $ppre_\mathrm{u}$ = $\mathrm{\{\langle {skills},~ matlab \rangle\}}$
\item [] $ppre_\mathrm{G_1}$ =  $ppre_\mathrm{G_2}$ = $ppre_\mathrm{G_5}$ = $\mathrm{\{\langle {skills},~ python \rangle\}}$
\item [] $ppre_\mathrm{G_3}$ = $\mathrm{\{\}}$
\end{itemize}

\noindent
The negation conjuncts are calculated as:
\begin{itemize}[leftmargin= 0cm]
\item [] $npre_\mathrm{u}$ = $\mathrm{\{\}}$, $npre_\mathrm{G_1}$ = $npre_\mathrm{G_2}$ = $npre_\mathrm{G_3}$ = $\mathrm{\{\langle {roomAcc},~ 2.04 \rangle\}}$
\item [] $npre_\mathrm{G_5}$ = $\mathrm{\{\}}$
\end{itemize}
These negation values are not present in user or all of its effective groups. Therefore, the algorithm creates directed graph (line 22-30) with vertices $V$ := $\mathrm{\{\langle {skills},~ matlab \rangle\}}$, $\mathrm{\{\langle {college},~ BUS \rangle\}}$. Edge in $E$ is drawn from $\mathrm{\{\langle {skills},~ matlab \rangle\}}$ to $\mathrm{\{\langle {college},~ BUS \rangle\}}$ as $\mathrm{\{\langle {skills},~ matlab \rangle\}}$ is a precondition conjunct in rule to add $\mathrm{\{\langle {college},~ BUS \rangle\}}$. Line 31 calculates $valset$ which in this case is same as $V$. Since no cycle exists in the graph, topological sort is created. The final reachability plan to satisfy the query $q_3$ is $plan$ :=  $\mathrm{{\mathsf{assign}(DeptAdmin,~u,~G_5)}}$, $\mathrm{{\mathsf{assign}( DeptAdmin,~u,~G_3)}}$, $\mathsf{add}$( DeptAdmin, u, skills, matlab), $\mathsf{add}$( BuildAdmin, u, college, BUS). The administrative requests must be executed as ordered in the reachability plan.

\section{Conclusion and Future Work}
\label{s:conc}
Attributes based access control defines permissions of entities based on their attributes. In this work, we presented reachability analysis for effective attributes of the user based on the direct attributes assignment to the user or its member user groups. We first stated the HGABAC model and \GURAG administrative model to provide some background. We defined a restricted form of \GURAG, referred as \miniGURANG and classified three schemes \miniGURANGZ, \miniGURANGO and \miniGURANGT to discuss different reachability solutions. In general, we proved the reachability problem for \rGURAG scheme is intractable as \ps-complete but with certain restrictions, polynomial time algorithms can also be achieved. For future works, we can develope more polynomial algorithms for some restricted forms and perform reachability analysis on other types of queries like effective user groups or minimum number of administrative requests to satisfy query.
\bibliographystyle{IEEEtran}
\bibliography{mybib-short}

\begin{IEEEbiographynophoto}{Maanak Gupta}
is an Assistant Professor in Computer Science at Tennessee Technological University, Cookeville, USA. He received M.S. and Ph.D. in Computer Science from the University of Texas at San Antonio (UTSA) and has also worked as a postdoctoral fellow at the Institute for Cyber Security (ICS) at UTSA. His primary area of research includes security and privacy in cyber space focused in studying foundational aspects of access control and their application in technologies including cyber physical systems, cloud computing, IoT and Big Data. He has worked in developing novel security mechanisms, models and architectures for next generation smart cars, smart cities, intelligent transportation systems and smart farming. He holds a B.Tech degree in Computer Science and Engineering from Kuruskhetra University, India, and an M.S. in Information Systems from Northeastern University, Boston.

\end{IEEEbiographynophoto}


\begin{IEEEbiographynophoto}{Ravi Sandhu}
is the founding Executive Director and Chief Scientist at the Institute for Cyber Security at the University of Texas at San Antonio, TX, where he holds the Lutcher Brown Endowed Chair in Cyber Security. He is a fellow of the ACM, IEEE and AAAS and an inventor on 30 patents. He was the past Editor-in-Chief of the IEEE Transactions on Dependable and Secure Computing, past founding Editor-in-Chief of the ACM Transactions on Information and System Security and a past Chair of ACM SIGSAC. He founded ACM CCS, SACMAT and CODASPY, and has been a leader in numerous other security conferences. His research has focused on security models and architectures, including the seminal role-based access control model. His papers have accumulated over 43,000 Google Scholar citations, including over 9,000 citations for the RBAC96 paper.
\end{IEEEbiographynophoto}
\setcounter{secnumdepth}{0}
\section{Appendix}
\textbf{Proof for Lemma 1}: This proof is an extension to proof discussed in \cite{jin2013reachability}. A Non-deterministic Turing machine can be used to implement following algorithm for each \rGURAG problem instance. Each state of the machine stores information to determine next possible states it can enter. In \rGURAG schemes, this information consists of current direct user attribute assignments, direct group attribute assignments, user to group assignments, attribute scopes, administrative rules, user groups and reachability query. The administrative rules are applied against current user or group attributes, or user to group assignments to get all next possible states. In each future state, Turing machine checks against the reachability query, and determines if the query is satisfied. If in a state the query is satisfied, Turning machine comes to halt or otherwise, same process repeats till a satisfied state is reached or it is concluded that the query is non-satisfiable. The size of each state is bounded by input to the state. It is understandable that polynomial amount of space is required to store information required in each state of Non-deterministic Turing machine. Hence, each problem scheme in Figure \ref{fig:rgurag} is in NPSPACE and therefore in PSPACE using Savitch's theorem \cite{savitch1970relationships}.

\smallskip
\noindent
\textbf{Discussion on Corollary 1}:
This section provides a brief overview of the analysis done in \cite{jin2013reachability} for \miniGURANZ~scheme. 

The analysis result for \RPSUPER query in \miniGURANZ~scheme is derived by reducing from role reachability problem. This role reachability problem for miniARBAC97 is proven \ps-complete \cite{sasturkar2006policy}, which by reduction makes \miniGURANZ~scheme as \ps-hard and using the same Lemma as \ref{lemma:pspace}, \miniGURANZ~scheme is \ps-complete.
This reduction uses role as one of the many attributes and map the administrative rules of miniARBAC97 to the corresponding rules in \miniGURANZ~as their expressive power is same. Our \miniGURANGZN~scheme extends \miniGURANZ~by introducing the notion of user groups and there corresponding administrative rules. Since, without user groups \miniGURANZ~and \miniGURANGZN~are same, the results for \RPSUPER in \miniGURANZ~still hold true and provide lower bound analysis. Therefore, we conclude that \RPSUPER for [\miniGURANGZN] is \ps-hard and using Lemma \ref{lemma:pspace}, it is in \ps-complete.

In GURA reachability \cite{jin2013reachability}, \RPSAME results for \miniGURANZ~uses reduction from SAS planning problem \cite{backstrom1995complexity} in artificial intelligence. Each state variable in [SAS, U, B] problem (proved to be \ps-complete in \cite{backstrom1995complexity}) is mapped to one value in scope of attribute $att$. The operators which update the state variables to true or false are mapped to administrative rules. This reduction is polynomial time which results \RPSAME for \miniGURANZ~in \ps-complete. Same results can be extended for \miniGURANGZN, where the values in scope for $att$ will be for the user and each user-group. The operators will be mapped to administrative rules for user and each user-group. Also each operator will change only one variable holding [SAS, U, B] restrictions and the reduction is in polynomial time. Henceforth,  \miniGURANGZN is \ps-hard problem and therefore \ps-complete using Lemma \ref{lemma:pspace}.

\end{document}